\documentclass[useAMS, usenatbib, letterpaper]{mn2e}

\usepackage{graphicx, bm, amssymb, xcolor}
\usepackage{verbatim}
\usepackage{threeparttable}
\usepackage[backref,breaklinks,colorlinks,citecolor=blue]{hyperref}
\usepackage[all]{hypcap}
\usepackage{times}

\newcommand{\noopsort}[1]{}

\defcitealias{krumholz15d}{Paper 1}

\newcommand{\vader}{\texttt{VADER}}

\begin{document}

\title[Star Formation in Galactic Centres]{A Dynamical Model for Gas Flows, Star Formation, and Nuclear Winds in Galactic Centres}

\author[Krumholz, Kruijssen, \& Crocker]{Mark R. Krumholz$^1$\thanks{mark.krumholz@anu.edu.au},
J.~M.~Diederik Kruijssen$^2$\thanks{kruijssen@uni-heidelberg.de}, and Roland M.~Crocker$^1$\thanks{roland.crocker@anu.edu.au}
\\ \\
$^1$Research School of Astronomy \& Astrophysics, Australian National University, Canberra, ACT 2611, Australia\\
$^2$Astronomisches Rechen-Institut, Zentrum f\"{u}r Astronomie der Universit\"{a}t Heidelberg, M\"{o}nchhofstra\ss e 12-14, D-69120 Heidelberg, Germany \\
}

\date{\today}

\pagerange{\pageref{firstpage}--\pageref{lastpage}} \pubyear{2016}

\maketitle

\label{firstpage}

\begin{abstract} 
We present a dynamical model for gas transport, star formation, and winds in the nuclear regions of galaxies, focusing on the Milky Way's Central Molecular Zone (CMZ). In our model angular momentum and mass are transported by a combination of gravitational and bar-driven acoustic instabilities. In gravitationally-unstable regions the gas can form stars, and the resulting feedback drives both turbulence and a wind that ejects mass from the CMZ. We show that the CMZ is in a quasi-steady state where mass deposited at large radii by the bar is transported inward to a star-forming, ring-shaped region at $\sim 100$ pc from the Galactic Centre, where the shear reaches a minimum. This ring undergoes episodic starbursts, with bursts lasting $\sim 5-10$ Myr occurring at $\sim 20-40$ Myr intervals. During quiescence the gas in the ring is not fully cleared, but is driven out of a self-gravitating state by the momentum injected by expanding supernova remnants. Starbursts also drive a wind off the star-forming ring, with a time-averaged mass flux comparable to the star formation rate. We show that our model agrees well with the observed properties of the CMZ, and places it near a star formation minimum within the evolutionary cycle. We argue that such cycles of bursty star formation and winds should be ubiquitous in the nuclei of barred spiral galaxies, and show that the resulting distribution of galactic nuclei on the Kennicutt-Schmidt relation is in good agreement with that observed in nearby galaxies.
\vspace{0.3in}
\end{abstract}
\begin{keywords}
Galaxy: nucleus --- galaxies: nuclei --- galaxies: spiral --- galaxies: star formation --- ISM: kinematics and dynamics  --- stars: formation 
\end{keywords}

\section{Introduction}
\label{sec:intro}

In the past decade it has become clear from observations that star formation in the Central Molecular Zone (CMZ; \citealt{morris96a}) of the Milky Way, and in the centres of other nearby galaxies \citep[e.g.,][]{barth95a,jogee02a}, deviates from the patterns of star formation and gas distribution that are observed at larger galactic radii. In the bulk of galactic discs, including that of the Milky Way, the molecular gas that fuels star formation is organised into clouds that are arranged in spiral patterns, either flocculent or grand design. In contrast, in the Milky Way's CMZ much of the gas is collected into a partially filled, ring-like stream of material $\sim 100$ pc from the Galactic Centre, which appears to be a persistent structure \citep{sofue95a, molinari11a, kruijssen15a, henshaw16a}. Clouds exist within the ring, but appear to form a well-defined time sequence in terms of their level of star formation activity \citep{longmore13b}. While rings such as this are occasionally seen at larger galactocentric radii (e.g., Andromeda), they are far from the typical arrangement of gas.

Second, the molecular gas in the bulk of spiral galaxies appears to form stars with a fairly constant depletion time (defined as the ratio of the gas surface density to the star formation surface density), with either no dependence \citep[e.g.,][]{bigiel08a, leroy08a, leroy13a} or only a weak dependence \citep[e.g.,][]{meidt13a, suwannajak14a} on the large scale rate of shear or other galactic-scale dynamics. In contrast, galactic centres exhibit a much wider range of depletion times than do the outer parts of discs \citep[e.g.,][]{saintonge12a, leroy13a, longmore13a}. Furthermore, in outer discs there is no obvious evidence for dynamical effects at all if one considers gas much denser than the $\sim 100$ cm$^{-3}$ traced by CO emission \citep[e.g.,][]{gao04b, garcia-burillo12a, usero15a}.  In contrast, given its budget of dense gas, the Milky Way's CMZ appears to be forming significantly fewer stars than one would expect if it had the same depletion time observed elsewhere. The present-day star formation rate of the CMZ is $\sim 0.05$ $M_\odot$ yr$^{-1}$ \citep{crocker12a, longmore13a, koepferl15a}, 
whereas the expected rate if the dense gas in the CMZ formed stars on a timescale similar to that found elsewhere in galaxies would be at least an order of magnitude larger. At the extreme end of this variation are CMZ objects such as ``The Brick" \citep{longmore12a, kauffmann13a, rathborne14a, rathborne15a, mills15a}, large clouds of extremely dense molecular gas that, if found in the outer Galaxy, would be expected to be intensely star-forming, yet in fact display almost no star formation activity.

A third potentially odd feature of Galactic Centre star formation is its burstiness. While star formation is always bursty when measured on sufficiently small scales simply as a result of finite molecular cloud masses and lifetimes \citep[e.g.][]{da-silva14b, kruijssen14c}, there is substantial evidence that the Milky Way's CMZ is significantly more episodic than the rest of the disc. Lines of evidence for episodic star formation in the CMZ include both direct star counts \citep{yusef-zadeh09a} that reveal more stars than would be expected given the present day production rate, and the presence of large off-plane bubbles  \citep{sofue84a, bland-hawthorn03a, su10a} that would appear to require $\sim 0.1$ $M_\odot$ yr$^{-1}$ to drive, somewhat higher than the present-day star formation rate, but not higher than the time-averaged star formation rate that would be inferred from the present-day mass of the stellar bulge  \citep{crocker12a, kruijssen14b, crocker15a}.

In \citet[hereafter \citetalias{krumholz15d}]{krumholz15d}, we introduced a model to explain some of the major observed features of the Milky Way CMZ and, by extension, the analogous regions of other barred spiral galaxies. The central idea of this model was to note that the Galactic Bar will transport a relatively continuous supply of gas from the inner Lindblad resonance (ILR; $r\sim 1$ kpc) to the outskirts of the CMZ disc ($r<500$ pc; \citealt{binney91a, kormendy04a, sormani15a}). Once deposited there, gas in the CMZ will be subject to periodic perturbations from the bar, which inside the ILR can drive acoustic instabilities that will simultaneously transport mass inward and pump up the gas velocity dispersion \citep{bertin89a, montenegro99a}, thereby preventing it from becoming self-gravitating and forming stars. This process will continue until the gas reaches $\sim 100$ pc, where the observed rotation curve of the Milky Way begins to turn over from flat to solid body, and the rate of shear drops. The loss of shear suppresses acoustic instabilities (which only occur when shear is present) and causes gas to accumulate until it becomes self-gravitating and star formation begins. (Indeed, the idea that low-shear regions tend to accumulate gas and produce rings goes back considerably before our model, e.g., \citealt{icke79b} and \citealt{fukunaga83a}.) We showed that this mechanism naturally produces the observed ring-like structure and explains its location, and that it naturally explains the long depletion times observed in in the CMZ. We further conjectured that, once star formation begins, stellar feedback would then expel much of the gas, leading to quenching until the bar replenished the gas supply, and explaining why star formation occurs in bursts, though we modify this picture in this paper.

While this model has a number of attractive features, the last portion of it necessarily remained conjectural, because we did not model the process of star formation feedback and gas ejection directly in \citetalias{krumholz15d}. We could not directly estimate the time scale of the bursts, for example, nor could we compute their magnitude, the partition of inflowing material between star formation and loss in a wind, and the level of variation we expect in the gas mass as a result of starbursts. In this paper we seek to remedy this situation by extending the model presented in \citetalias{krumholz15d} with a treatment of star formation feedback and wind ejection. As in \citetalias{krumholz15d}, we focus first on the Milky Way's CMZ, because that is the region for which we have by far the best dynamical information, but we then extend the model to other galaxies.

The plan for the remainder of this paper is as follows. in \autoref{sec:model} we present our basic model, and highlight the new treatment of star formation and feedback that we have added in comparison to \citetalias{krumholz15d}. In \autoref{sec:results} we present simulation results. We discuss the implications of these results in \autoref{sec:discussion}, and summarise and discuss prospects for future work in \autoref{sec:conclusion}.

\section{Model}
\label{sec:model}

The model we build for the Milky Way's Central Molecular Zone (CMZ) is a generalisation of the one presented in \citetalias{krumholz15d}. Here we summarise the most salient aspects of that model, referring readers to \citetalias{krumholz15d} for full details, before moving on to the new aspects of the model included here. Unless otherwise noted, all parameter choices made in this paper are identical to the fiducial ones made in \citetalias{krumholz15d}. All the simulation code used for this project is publicly available from \url{https://bitbucket.org/krumholz/cmzsf}.

\subsection{Dynamical Evolution}

We approximate the gas in the CMZ as an axisymmetric thin disc characterised by a surface density $\Sigma$ and velocity dispersion $\sigma$, both as a function of radius $r$ from the Galactic Centre. The gas orbits in a potential derived from the measurements of \citet{launhardt02a}. We use these measurements to produce a smooth, interpolated rotation curve $v_\phi(r)$ from which we can derive the dimensionless index $\beta = d\ln v_\phi/d\ln r$ that describes the rate of shear; formally, the dimensionless shear rate is $1-\beta$. We treat the rotation curve as constant in time.\footnote{Our approximation that the rotation curve is constant limits the total time for which we can run our simulations to be such that the mass of stars formed during the simulation is small compared to the dynamical mass responsible for producing the rotation curve. For the run duration of 500 Myr that we adopt below, this condition is satisfied for all our runs; in our fiducial case the mass added to the domain is below 10\% of the dynamical mass interior to the radius where stars form, and for all runs it is below 20\%.}
We evolve the gas using the \vader\ code of \citet{krumholz15a}, which solves the equations of mass, energy, and angular momentum conservation for the disc in conservative form. As in \citetalias{krumholz15d}, we place the inner and outer edges of the region to be simulated at $r = 10$ and $450$ pc, respectively, and use 512 computational zones uniformly spaced in $\log r$. Our model here differs from that in \citetalias{krumholz15d} only in that we include source terms in the equations to represent the effects of star formation and winds. Formally, the equations we solve are
\begin{eqnarray}
\frac{\partial}{\partial t}\Sigma + \frac{1}{r} \frac{\partial}{\partial r} \left(r v_r \Sigma\right) & = & -\dot{\Sigma}_* - \dot{\Sigma}_{\rm wind} \\
\lefteqn{\frac{\partial}{\partial t} E + \frac{1}{r} \frac{\partial}{\partial r} \left[rv_r(E+P)\right] - \frac{1}{r}\frac{\partial}{\partial r} \left(r \frac{v_\phi \mathcal{T}}{2\pi r^2}\right)}
\qquad\qquad\qquad\qquad
\nonumber \\
 & = & \dot{E}_{\rm SF,turb} - \dot{E}_{\rm rad},
\end{eqnarray}
where the source terms on the right-hand side of the first equation represent the rates of change of gas surface density due to star formation and loss by winds, while those in the second equation represent the rate of change of turbulent energy due to star formation feedback and due to radiative losses from shocks. We discuss the values of these terms below. In these equations $P = \Sigma \sigma^2$ is the vertically-integrated pressure, $v_r$ is the radial velocity, and $\mathcal{T}$ is the turbulent torque, which is related to $v_r$ via angular momentum conservation:
\begin{equation}
v_r = \frac{\partial\mathcal{T}/\partial r}{2\pi r \Sigma v_\phi (1+\beta)}.
\end{equation}

A key parameter of this model is the dimensionless rate of angular momentum transport $\alpha$ produced by instabilities, which determines $\mathcal{T}$ via
\begin{equation}
\mathcal{T} = -2\pi r^2 \alpha P(1-\beta).
\end{equation}
As in \citetalias{krumholz15d}, we consider two sources of transport: gravitational and acoustic instability. The former is parameterised by the usual \citet{toomre64a} $Q$ parameter. The latter instability can occur when gas orbits inside the inner Lindblad resonance of a periodic perturber, in this case the Galactic Bar. It arises when pressure waves within the disc driven by the bar cause the perturbed gas orbits to align, leading to a growing mode. The instability grows most strongly in regions of weak self-gravity and high shear. Both gravitational and acoustic modes can be combined into a single dispersion relation, derived by \citet{montenegro99a}. In our simulations, we obtain numerical solutions to this dispersion relation at each radius. When an unstable mode is present, we compute the growth timescale $t_{\rm growth}$ of the fastest growing mode. In unstable regions, we take
\begin{equation}
\alpha = \min(\alpha_0 e^{1-t_{\rm growth}/t_{\rm orb}}, 1),
\end{equation}
where $t_{\rm orb}$ is the local orbital period in the disk. We adopt the same fiducial value $\alpha_0 = 1$ as in \citetalias{krumholz15d}, so that, in regions of the disk where an unstable mode has a growth timescale equal to or smaller than the orbital period, the rate of transport corresponds to a large value $\alpha \approx 1$. We argue in \citetalias{krumholz15d} that, given the nature of the instabilities we are considering, this is the most plausible value.

A second key parameter in our models is the rate of radiative losses from the disc, $\dot{E}_{\rm rad}$. These losses occur due to radiative shocks produced by the turbulence in the disc, and result in the full disc energy being radiated away each dynamical time. We compute $\dot{E}_{\rm rad}$ exactly as in \citetalias{krumholz15d}. We pause here to note an important implication of the value of $\dot{E}_{\rm rad}$: the loss of turbulent energy on a flow crossing timescale tends to push galactic discs toward $\alpha \approx 1$ in non-star-forming regions. The reason is that, in the absence of star formation as an energy source, maintaining energy balance in a galactic disc requires that the rate of energy release by inward transport of mass balance the rate of energy dissipation. If the timescale for energy dissipation is a dynamical time, then the rate of inward mass flow required for balance corresponds to $\alpha\approx 1$, with the exact value depending on the exact energy dissipation rate, the gas fraction, and the rotation curve \citep{krumholz10c}.

The disc simulation requires boundary conditions at the inner and outer edges. For the inner boundary, we set the mass flux to be zero; in practice we find that no significant amount of mass approaches the inner boundary, so this choice has no practical effect. At the outer boundary, we  impose a fixed inward mass flux $\dot{M}_0$, for which we consider a range of possible values. This mass flux is provided by material that is removed from its circular orbit by the Galactic bar and transported inwards to settle into the CMZ \citep{binney91a, kormendy04a, crocker12a, sormani15a}. The mass transport rate is uncertain, but observations suggest it lies in the range $\dot{M}_{\rm in} = 0.1 - 1.0$ $M_\odot$ yr$^{-1}$, so we consider this range in our work. We set the velocity dispersion of this inward-flowing material to $\sigma_{\rm in} = 40$ km s$^{-1}$, following \citetalias{krumholz15d}. We initialise all our simulations by placing a uniform surface density of $0.01$ $M_\odot$ pc$^{-2}$ with a velocity dispersion of $40$ km s$^{-1}$ in all zones, thereby beginning the simulations in a nearly gas-free state.

\subsection{Star Formation}

Where our model differs from that of \citetalias{krumholz15d} is that we have added models for star formation and feedback, which were absent from that paper. To determine where star formation will occur, we must answer the question of where the gas becomes self-gravitating. Let $H_g$ be the gas scale height, which we compute from the gas surface density, velocity dispersion, and stellar density as in \citetalias{krumholz15d}. Formally we can write the rate of star formation per unit area in the disc as
\begin{equation}
\dot{\Sigma}_* = \epsilon_{\rm ff} \frac{\Sigma}{t_{\rm ff}}
\end{equation}
where
\begin{equation}
t_{\rm ff} = \sqrt{\frac{3\pi H_g}{16 G \Sigma}}
\end{equation}
is the free-fall time at the mid-plane (using $\Sigma/2H_g$ as the gas density), and $\epsilon_{\rm ff}$ is the dimensionless star formation rate per free-fall time \citep{krumholz14c, padoan14a}. The value of $\epsilon_{\rm ff}$ depends on the degree of gravitational boundedness as characterised by the virial ratio $\alpha_{\rm vir}$, and also on the Mach number, plasma $\beta$, compressive to solenoidal ratio of the turbulence \citep[e.g.][]{krumholz05c, padoan11a, federrath12a, federrath13a}. However, $\alpha_{\rm vir}$, is by far the most important parameter, and is the only one we can easily calculate given our simple model. To determine its value, we note that the midplane pressure in our disc is
\begin{equation}
p_{\rm mp} = \frac{\Sigma\sigma^2}{H_g}.
\end{equation}
For a disc supported by pressure against self-gravity, we have \citep[e.g.,][]{krumholz05c}
\begin{equation}
\label{eq:pmp_eq}
p_{\rm mp,eq} = \frac{\pi}{2} G \Sigma^2.
\end{equation}
Note that we have $\Sigma^2$ rather than $\Sigma (\Sigma + \rho_* H_g)$ here because we are interested in the support of the gas against its own self-gravity, discounting the contribution from the gravity of the stars. From these two expression, we can express the virial parameter of the gas as
\begin{equation}
\alpha_{\rm vir} = \frac{p_{\rm mp}}{p_{\rm mp,eq}},
\end{equation}
so that gas becomes self-gravitating as $\alpha_{\rm vir} \rightarrow 1$ from above, and is non-self-gravitating if $\alpha_{\rm vir} \gg 1$. Note that, because we calculate the scale height under the assumption of hydrostatic equilibrium (see \citetalias{krumholz15d}), our model does not permit $\alpha_{\rm vir} < 1$, since $\alpha_{\rm vir} < 1$ can be achieved only under non-equilibrium conditions.
Given a value $\alpha_{\rm vir}$, we determine $\epsilon_{\rm ff}$ using an approximation suggested by \citet{padoan11a}, which is that $\epsilon_{\rm ff}$ declines approximately exponentially with $\alpha_{\rm vir}$. Both observations and simulations suggest that $\epsilon_{\rm ff} \sim 0.01$ for $\alpha_{\rm vir}\approx 1$ (see the reviews by \citealt{krumholz14c} and \citealt{padoan14a}, and references therein), and we expect that $\epsilon_{\rm ff} \rightarrow 1$ as $\alpha_{\rm vir} \rightarrow 0$. Thus we adopt the relationship
\begin{equation}
\epsilon_{\rm ff} = \exp\left[\alpha_{\rm vir} \log(\epsilon_{\rm ff,0})\right],
\end{equation}
with $\epsilon_{\rm ff,0} = 0.01$ as a fiducial choice. This expression has all the properties we desire: $\epsilon_{\rm ff} \rightarrow 1$ as $\alpha_{\rm vir} \rightarrow 0$, $\epsilon_{\rm ff} = 0.01$ at $\alpha_{\rm vir} = 1$, and $\epsilon_{\rm ff}$ declines exponentially as $\alpha_{\rm vir}$ rises. While the value of $\epsilon_{\rm ff}$ is tightly constrained by observations to lie near our fiducial choice \citep[e.g.,][]{krumholz07e, krumholz12a, federrath13c, krumholz14c, evans14a, salim15a, heyer16a}, we also consider the effects of varying $\epsilon_{\rm ff,0}$.

Before moving on we note that, although we have phrased our star formation rate as a function of $\alpha_{\rm vir}$, the virial ratio in our models is closely related to the Toomre $Q$ of the gas. One can show that $\alpha_{\rm vir} \approx 1$ is equivalent to $Q\approx 1$, and thus one may view the dependence of $\epsilon_{\rm ff}$ on  $\alpha_{\rm vir}$ in our model as qualitatively equivalent to the condition that star formation starts up as $Q$ approaches 1.

\subsection{Stellar Feedback}
\label{ssec:feedback}

Feedback from stars in our model takes two forms: injection of energy and ejection of mass in the form of winds. Both processes are governed by the momentum input of massive stars, since stellar winds and supernova ejecta that interact with the dense gas in the CMZ will become radiative very quickly, a point to which we will return in \autoref{ssec:windprop}. The first step in our model of feedback is therefore to compute the momentum injection rate from star formation. To do so, we use \texttt{starburst99} \citep{leitherer99a, vazquez05a} to compute the type II supernova rate per unit mass $\Gamma_{\rm SN}(t)$, the bolometric luminosity per unit mass $\mathcal{L}(t)$, and the wind momentum injection rate per unit mass $\mathcal{P}_{\rm wind}(t)$ for simple stellar populations of age $t$ with a \citet{kroupa02c} IMF. The starlight carries a momentum per unit stellar mass per unit time $L(t)/c$. For the supernovae, we adopt a momentum injection per supernova of $p_{\rm SN} = 3\times 10^5$ $M_\odot$ km s$^{-1}$ based on recent simulations \citep[e.g.,][]{martizzi15a, kim15a, walch15b, gentry16a}, giving a supernova momentum injection rate $\Gamma_{\rm SN}(t) p_{\rm SN}$.\footnote{One might worry that the momentum budget would be smaller at the $n\sim 10^4$ cm$^{-3}$ densities found in the CMZ than for the $n\sim 1-100$ cm$^{-3}$ densities found at larger radii, because supernova remnants would become radiative more quickly. However, the simulations show that supernova momentum budget is not very sensitive to density, with fits to the simulation results giving scalings that vary from $p_{\rm SN} \propto n^{-0.06}$ \citep{gentry16a} to $p_{\rm SN} \propto n^{-0.19}$ \citep{martizzi15a}. Moreover, the clustering expected in high density regions can also enhance the momentum budget by a factor of several, pushing in the other direction \citep{gentry16a}. Thus our fiducial estimate should be reasonable even in the CMZ.} The total momentum injection rate per unit time per unit area in our simulations is then simply the sum of these three quantities, convolved with the star formation history, i.e.,
\begin{eqnarray}
\lefteqn{\frac{d\dot{p}}{dA}(t) = } \nonumber \\
& & \int_0^t \dot{\Sigma}_*(t - t') \left[p_{\rm SN} \Gamma_{\rm SN}(t') + \frac{\mathcal{L}(t')}{c} + \mathcal{P}_{\rm wind}(t')\right]\, dt'.
\end{eqnarray}
Since we know the star formation history from the prescription above, this quantity is straightforward to evaluate.

We pause here for three brief comments on the model. First, although we have included winds, radiation pressure, and supernovae, our choice of $p_{\rm SN}$ implies that supernovae are by far the most important form of feedback; winds and radiation pressure are small perturbations on this. Second, we consider only star formation feedback and gravity as sources of turbulence, which means that we are omitting a potential contribution to turbulence from a galactic fountain or from accretion directly onto the CMZ from above (rather than through the disk). These effects could conceivably increase the turbulent velocity dispersion from what we find, but are very poorly constrained either observationally or theoretically. Third, note that we have not included a contribution from trapped infrared radiation pressure. The significance of such an effect has been subject to extensive discussion in the literature in the past few years \citep[e.g.,][]{krumholz09d, murray10a, krumholz12c, krumholz13a, davis14a, rosdahl15a, tsang15a}. We will not rehash that discussion here, but we note that, even in the simulations where trapped infrared radiation pressure is found to be most effective, it becomes significant only when the gas column density and luminosity are so high that the gas disc is optically thick even for radiation whose colour temperature is equal to that of the dust photosphere; \citet{krumholz13a} show that the condition is met only when the gas surface density exceeds $\sim 5000$ $M_\odot$ pc$^{-2}$ and the star formation surface density exceeds $\sim 1000$ $M_\odot$ pc$^{-2}$ Myr $^{-1}$. While such extreme combinations of gas and star formation surface density may exist on $\lesssim 1$ pc scales in Galactic Centre star-forming regions such as Sgr B2 \citep[e.g.,][]{schmiedeke16a}, they are never realised over the larger scales with which we are concerned, either in the real Galactic Centre or in our models. 

The second step is to consider where the momentum will be deposited. The simplest assumption would be to inject momentum where the stars form, but this ignores the fact that the stars will form with some velocity dispersion relative to the gas out of which they are born. Thereafter they are not constrained to move on exactly the same orbits as the gas. Since supernovae occur over timescale of $\sim 10$ Myr after star formation, and the orbital period at 100 pc from the Milky Way's centre is only $\sim 3$ Myr, stars that are on slightly different orbits than the gas from which they form will have time to drift some distance from their birth sites before exploding, and this will blur out the location where they deposit their momentum. We do not attempt to model this evolution in detail, and instead resort to parameterising it. Specifically, rather than compute the momentum injection rate using the true star formation rate $\dot{\Sigma}_*(r,t)$ in our simulation, we use the convolution of the star formation rate with a Gaussian blur,
\begin{equation}
\dot{\Sigma}_{*,\rm eff}(r,t) = N^{-1} \int \exp\left[-\frac{(r - r')^2}{2(\epsilon_r r')^2}\right] \dot{\Sigma}_*(r',t)\, dr',
\end{equation}
where the normalisation factor $N$ is set by the requirement that $\int\dot{\Sigma}_*\, dA = \int\dot{\Sigma}_{*,\rm eff}\, dA$, i.e., that the total amount of momentum injected, integrated over the area of the disc, remain unchanged. The dimensionless quantity $\epsilon_r$ parameterises the amount by which the stars spread out relative to the gas from which they form. Thus the rate of momentum injection in our simulations becomes
\begin{eqnarray}
\frac{d\dot{p}}{dA}(r,t) & = & \int_0^\infty \dot{\Sigma}_{*, \rm eff}(r, t - t') \cdot {}\nonumber \\
& & \; \left[p_{\rm SN} \Gamma_{\rm SN}(t') + \frac{\mathcal{L}(t')}{c} + \mathcal{P}_{\rm wind}(t')\right]\, dt'.
\end{eqnarray}

To decide on a fiducial value of $\epsilon_r$, note that if a population of stars begins on a circular orbit with radius $r$ and a velocity $v_\phi$, and their orbits are perturbed by a random velocity $v_*$, the resulting elliptical orbits will be confined to a range of radii \citep{binney87a}
\begin{equation}
r_* = r \left(1 \pm \frac{4}{3} \frac{v_*}{v_\phi}\right).
\end{equation}
Under the conditions observed in the CMZ, only some $\sim50$ per cent of all stars are expected to form in bound clusters \citep[e.g.][]{kruijssen12a,adamo15a}, with the rest forming in unbound associations. The unbound stars will drift apart at the internal velocity dispersions of the gas clouds from which they form \citep{efremov98a}, while the bound clusters will move together, dispersing from their birth sites at the overall centre of mass velocity of the cluster. We do not have direct measurements of either the velocity dispersions of associations or the bulk velocities of clusters, but we note that, at larger galactic radii, bound clusters and unbound associations appear to have roughly the same velocity dispersions, so we can use the measured velocity dispersions within CMZ clusters as a rough proxy for the typical velocity dispersion $v_*$. Observed one-dimensional velocity dispersions in Galactic centre star clusters such as the Arches \citep{clarkson12a} and Quintuplet \citep{stolte14a} are typically $\approx 5-6$ km s$^{-1}$, and certainly no more than 10 km s$^{-1}$, and these clusters are formed at $r\approx 90$ pc from the Galactic centre, where the circular velocity $v_\phi \approx 150$ km s$^{-1}$. This suggests that $(4/3)(v_*/v_\phi)\sim 0.05$, and so we adopt $\epsilon_r = 0.05$ as a fiducial value. We also explore variations around this choice.


With the rate of momentum injection in hand, we can now proceed to compute the rate at which star formation feedback both drives turbulence and launches winds. Following \citet{krumholz06d}, \citet{matzner07a}, \citet{goldbaum11a}, and \citet{faucher-giguere13a}, we approximate that supernova remnants and similar bubbles merge with the background turbulence and add their energy to it once their expansion velocity decreases to the turbulent velocity, in which case the rate of energy injection into turbulence produced by a momentum injection rate per unit area $d\dot{p}/dA$ is approximately
\begin{equation}
\left(\frac{d\dot{E}}{dA}\right)_{\rm SF,turb} = \sigma\left(\frac{d\dot{p}}{dA}\right).
\end{equation}
To avoid producing unphysically-large velocity dispersions in very low surface density cells, we suppress energy injection in cells with surface densities below a minimum value of $1$ $M_\odot$ pc$^{-2}$. Thus our final expression for the rate of energy injection by star formation is
\begin{equation}
\label{eq:edotturb}
\left(\frac{d\dot{E}}{dA}\right)_{\rm SF,turb} = \sigma\left(\frac{d\dot{p}}{dA}\right) e^{-\Sigma_{\rm lim}/\Sigma},
\end{equation}
with $\Sigma_{\rm lim} = 1$ $M_\odot$ pc$^{-2}$.

The final term we must compute is the rate at which momentum injection drives winds off the disc. We compute this rate following the formalism of \citet{thompson16a}. The essential idea of this model is that turbulence will produce a lognormal distribution of column densities in the disc. For a fixed rate of momentum injection per unit area, one can compute a critical column density below which the inertia of the gas is small enough that the upwards momentum injection produces a force that exceeds the force of gravity, leading material to be ejected. The rate of mass ejection depends on the ratio of the momentum injection rate to the mean Eddington injection rate, and on the Mach number of the turbulence, which determines the dispersion of column densities. The Mach number is simply $\mathcal{M} = \sigma/\sigma_{\rm th}$ where $\sigma_{\rm th}$ is the thermal velocity dispersion, which we take to be $0.5$ km s$^{-1}$ as in \citetalias{krumholz15d}.

To compute the Eddington injection rate, we must know the depth of the potential from which the gas must escape, including both the gaseous and stellar\footnote{``Stellar" here should be understood to include any contribution from collisionless dark matter as well.} contributions. The former is easy to compute: for gas of surface density $\Sigma$ in a thin disc, the gravitational acceleration is simply $g_{\rm gas} = 2\pi G \Sigma$, independent of height. The corresponding acceleration from the stellar potential is somewhat trickier to estimate, because the stars have a much larger scale height than the gas, and thus the gravitational acceleration experienced by a parcel of gas will increase as it rises above the midplane in a wind. To escape from the CMZ and not simply be puffed above the disc to fall back, the gas must have enough momentum to overcome the gravitational acceleration well above the disc. Computing this properly would require knowledge of the full three-dimensional stellar potential, which is only poorly constrained, but we can make a rough estimate. Following Paper I, we note that, in spherical symmetry, the stellar mass density at radius $r$ that is required to produce a rotation curve with velocity $v_\phi$ is given by
\begin{equation}
\rho_{*,\rm sphere} = (1+2\beta) \frac{v_\phi^2}{4\pi G r^2},
\end{equation}
and for such a spherical distribution the characteristic scale height is $\sim r$. We therefore approximate the stellar acceleration as $g_* \approx 2\pi G \rho_{*,\rm sphere} r$. A more flattened distribution would raise $\rho_*$ but decrease the scale height of the stellar distribution by the same factor, and thus produce about the same net result for the acceleration.

Combining the gaseous and stellar contributions, the Eddington momentum injection rate in a region with gas surface density
 $\Sigma$ is
 \begin{equation}
\label{eq:pdotedd}
\left(\frac{d\dot{p}}{dA}\right)_{\rm Edd} = \Sigma (g_{\rm gas} + g_*),
\end{equation}
and, following \citet{thompson16a}, we define the parameter $x_{\rm crit}$ as
\begin{equation}
\label{eq:xcrit}
x_{\rm crit} = \ln \left[\frac{d\dot{p}/dA}{\left(d\dot{p}/dA\right)_{\rm Edd}}\right].
\end{equation}
Note that we use $(g_{\rm gas} + g_*)$ rather than simply $g_{\rm gas}$ when computing the Eddington rate, as opposed to our approach in computing the virial ratio (c.f.~\autoref{eq:pmp_eq}), because for the latter we are concerned with whether self-gravity can induce the gas to collapse, while for the former we are concerned with the question of whether supernovae inject enough momentum into the gas to unbind it from both itself and from the stellar potential. Given $\mathcal{M}$ and $x_{\rm crit}$, \citet{thompson16a} show that the mass ejection rate is given by
\begin{equation}
\label{eq:sigmawind}
\dot{\Sigma}_{\rm wind} = \zeta \Sigma \frac{\sigma}{H_g},
\end{equation}
where
\begin{eqnarray}
\zeta & = & \frac{1}{2}\left[1 - \textrm{erf}\left(\frac{-2 x_{\rm crit} + \sigma_{\ln\Sigma}^2}{2\sqrt{2}\sigma_{\ln \Sigma}}\right)\right] \\
\sigma_{\ln\Sigma}^2 & = & \ln\left(1 + R\frac{\mathcal{M}^2}{4}\right) \\
R & = & 0.5 \left(\frac{\mathcal{M}^{-1.0}-1}{1-\mathcal{M}^{1.0}}\right).
\end{eqnarray}
Physically, \autoref{eq:sigmawind} simply asserts that gas with little enough inertia to be accelerated to the escape speed in a disc crossing time will be removed on that same timescale, while material of higher inertia, as implied by higher surface density, will not. Note that the dispersion in column densities $\sigma_{\ln\Sigma}$ is smaller than the corresponding dispersion in volume density for the same Mach number as a result of line of sight averaging. All the above expressions are valid in the limit $\mathcal{M}\gg 1$. 

As with energy injection, we exponentially suppress this effect once the surface density has been driven too low, in order to avoid generating unphysically low surface density cells that produce numerical problems. Thus in our code we modify \autoref{eq:sigmawind} to
\begin{equation}
\dot{\Sigma}_{\rm wind} = \zeta \Sigma \frac{\sigma}{H_g} e^{-\Sigma_{\rm lim}/\Sigma}.
\end{equation}

\subsection{Numerical Limits}

One final modification we make to the code is to impose a floor on the column density and a corresponding ceiling on the temperature. We do this because, after very long run times, cells near the inner edge of our grid can reach very low column densities and very high velocity dispersions not as a result of winds, but simply as a result of advection converting gravitational potential energy to velocity dispersion. This does not affect the results or the ability of the code to run, but it does result in time steps that are inconveniently small. We therefore add the following purely numerical source terms in all cells:
\begin{eqnarray}
\dot{\Sigma}_{\rm num} & = & \frac{\Sigma_{\rm floor}}{r/v_\phi} \left[\frac{e^{\Sigma_{\rm floor}/\Sigma}}{1 + e^{(\Sigma/\Sigma_{\rm floor})^2}}\right] \\
\dot{E}_{\rm num} & = & -\frac{\Sigma \sigma_{\rm NT}^2}{r/v_\phi}\left[\frac{e^{\sigma_{\rm NT}/\sigma_{\rm ceil}}}{1 + e^{(\sigma_{\rm ceil}/\sigma_{\rm NT})^2}}\right] 
\end{eqnarray}
where $\sigma_{\rm NT} = \sqrt{\sigma^2 - c_s^2}$ is the non-thermal velocity dispersion, $c_s = 0.5$ km s$^{-1}$ is our adopted thermal sound speed, $\Sigma_{\rm floor} = 10^{-4}$ $M_\odot$ pc$^{-2}$, and $\sigma_{\rm ceil} = 400$ km s$^{-1}$. Thus these terms artificially add mass and remove energy to keep the surface density from falling below $10^{-4}$ $M_\odot$ pc$^{-2}$ and the velocity dispersion from increasing above 400 km s$^{-1}$; both source terms are suppressed as $e^{-x^2}$ in cells not near these limits. We have verified that both of these source terms change the total mass or energy in the computational domain by only a tiny amount over the full course of the simulations, while increasing the mean time step by a factor of $\sim 100$.

\section{Results}
\label{sec:results}

\begin{table*}
\centering
\caption{List of simulations. \label{tab:sims}}
\begin{tabular}{@{}l@{\qquad}ccc@{\qquad}ccc@{}}
\hline
& \multicolumn{3}{c}{Input Parameters} & \multicolumn{3}{c}{Results}\\
Run Name & $\dot{M}_{\rm in}$& $\epsilon_r$ & $\epsilon_{\rm ff, 0}$ &
SFE & $\nu_{\rm max}^{-1}$ (observed) 
& $\nu_{\rm min}^{-1}$ (observed) \\
&  [$M_\odot$ yr$^{-1}$]  & & & & [Myr] & [Myr]
\\
\hline
m01r050f10   &  0.1   &   0.050   &   0.010   &   0.92   &   23 (23)   &   10 (10)  \\
m03r025f10   &  0.3   &   0.025   &   0.010   &   0.70   &   42 (42)   &    6 (8)  \\
m03r050f05   &  0.3   &   0.050   &   0.005   &   0.72   &   23 (23)   &   10 (10)  \\
m03r050f10   &  0.3   &   0.050   &   0.010   &   0.72   &   21 (21)   &    5 (8)  \\
m03r050f20   &  0.3   &   0.050   &   0.020   &   0.67   &   42 (42)   &    4 (8)  \\
m03r100f10   &  0.3   &   0.100   &   0.010   &   0.59   &   15 (15)   &    7 (7)  \\
m10r050f10   &  1.0   &   0.050   &   0.010   &   0.48   &   42 (42)   &    7 (8)  \\
\hline
\end{tabular}
\begin{tablenotes}
\item Here SFE is the star formation efficiency, defined as the time-averaged ratio of mass converted to stars to mass converted into stars plus lost to the wind (see \autoref{eq:SFE}). Note that this is distinct from both the instantaneous star formation efficiency or a single cloud and the star formation rate per free-fall time $\epsilon_{\rm ff}$. For the timescales $\nu_{\rm max}^{-1}$ and $\nu_{\rm min}^{-1}$, the first figure is the value computed using the true star formation rate, while the second (in parentheses) is the figure using the observationally-inferred star formation rate.
\end{tablenotes}
\end{table*}

In \autoref{tab:sims} we summarise the full set of simulations that we have run, and collect various quantitative results for them. Simulations vary only in the value of the accretion rate $\dot{M}_{\rm in}$ into the CMZ, the value of the parameter $\epsilon_r$ that determines the radial extent over which stellar feedback is spread, and the parameter $\epsilon_{\rm ff,0}$ that defines the rate of star formation per free-fall time at a virial ratio of unity; simulation names follow the convention m\textit{XX}r\textit{YYY}f\textit{ZZ}, where $XX=10\dot{M}_{\rm in}/(M_\odot\,\mathrm{yr}^{-1})$, $YYY=1000\epsilon_r$, and $ZZ=1000\epsilon_{\rm ff}$. All other parameters are as described in \autoref{sec:model}, or in \citetalias{krumholz15d}. We run all simulations for 500 Myr.

\subsection{Qualitative Behaviour}
\label{ssec:example}

\begin{figure*}
\includegraphics[width=\textwidth]{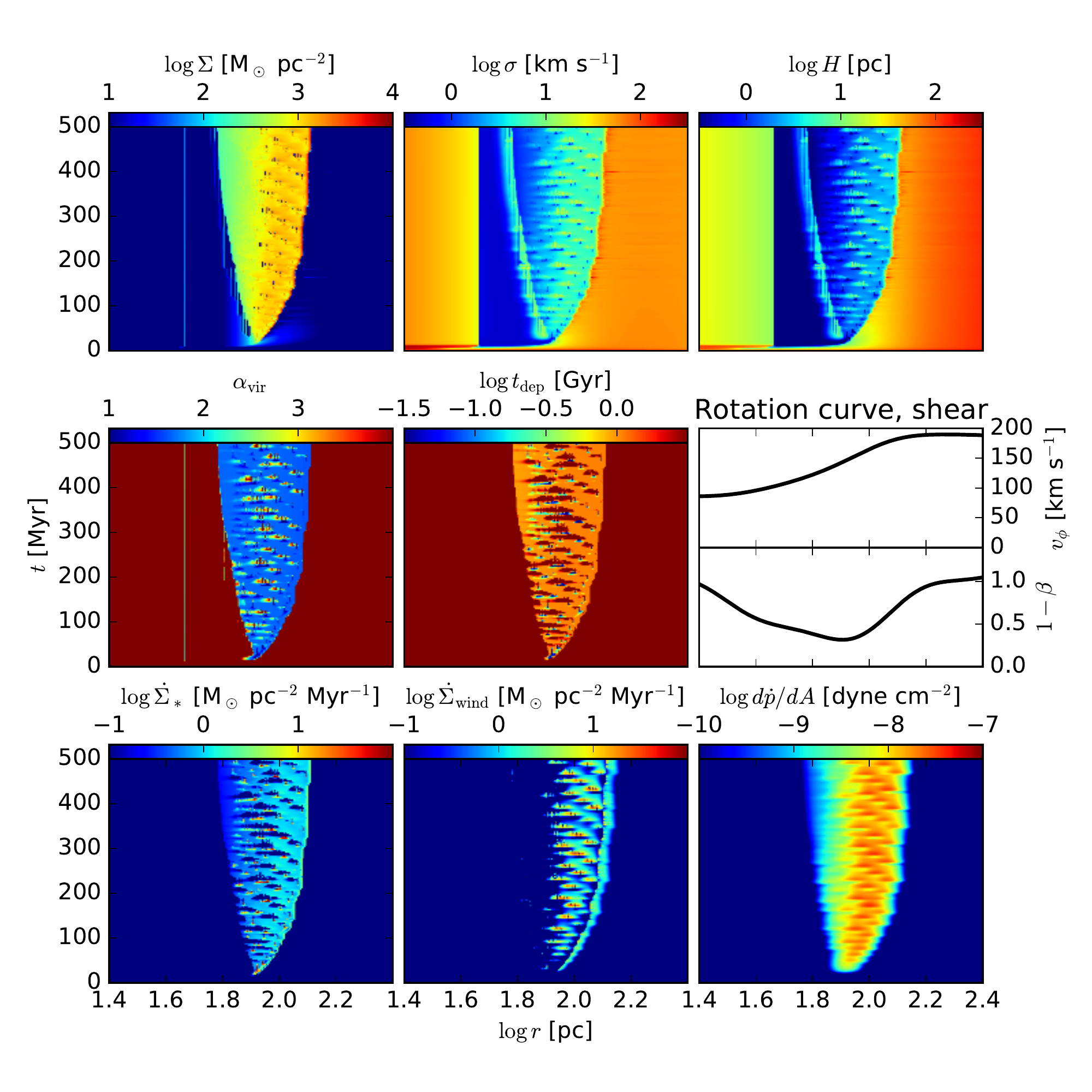}
\caption{
\label{fig:summary_fiducial}
Summary of the outcome of the fiducial simulation m03r050f10. In each panel, radial position is indicated on the $x$ axis and evolution time on the $y$ axis. Coloured panels indicate the values of the quantities indicated in the colour bars: gas surface density $\Sigma$, velocity dispersion $\sigma$, scale height $H$, virial ratio $\alpha_{\rm vir}$, depletion time $t_{\rm dep} = \Sigma/\dot{\Sigma}_*$, star formation rate $\dot{\Sigma}_*$, wind mass launching rate $\dot{\Sigma}_{\rm wind}$, and momentum injection rate $d\dot{p}/dA$. In the two centre-right panels, the line plots show the rotation velocity $v_\phi$ (top) and the  dimensionless rate of shear $1-\beta$ (bottom) as a function of radius. Note that we plot only a portion of the simulation domain in order to emphasise interesting features. 
}
\end{figure*}

We first focus on run m03r050f10 ($\dot{M}_{\rm in} = 0.3$ $M_\odot$ yr$^{-1}$, $\epsilon_r = 0.05$, $\epsilon_{\rm ff} = 0.01$), since it was run with our fiducial parameter choices, and many of the qualitative features we find in this run are common to all the simulations. \autoref{fig:summary_fiducial} summarises the outcome of this simulation. Gas enters from the outer edge of the computational domain and flows inward toward the origin as a result of acoustic instability.\footnote{We caution that this acoustic instability-dominated region is at the edge of applicability for our thin disc model. The transport equations that \texttt{VADER} solves are valid to order $(H_g/r)^2$ \citep{krumholz10c, krumholz15a}, and in the acoustic region $H_g/r$ is in the range $0.3-0.5$. Thus in this region we are dropping terms that are smaller than the ones we have retained by only $\approx 10-25\%$. That said, since there are no interesting dynamics in this region, and the gas simply flows through, the impact of such errors is likely to be minimal in any event.}
 Just inside $100$ pc, where the rotation curve turns from near-flat to near-solid body, this instability shuts off due to the loss of shear. The ``dead zone" where the shear is too small to drive acoustic instability is most easily visible in the plot of gas velocity dispersion, where it manifests as a region where the dispersion falls to low values until star formation begins and pumps it back up.

\begin{figure}
\includegraphics[width=\columnwidth]{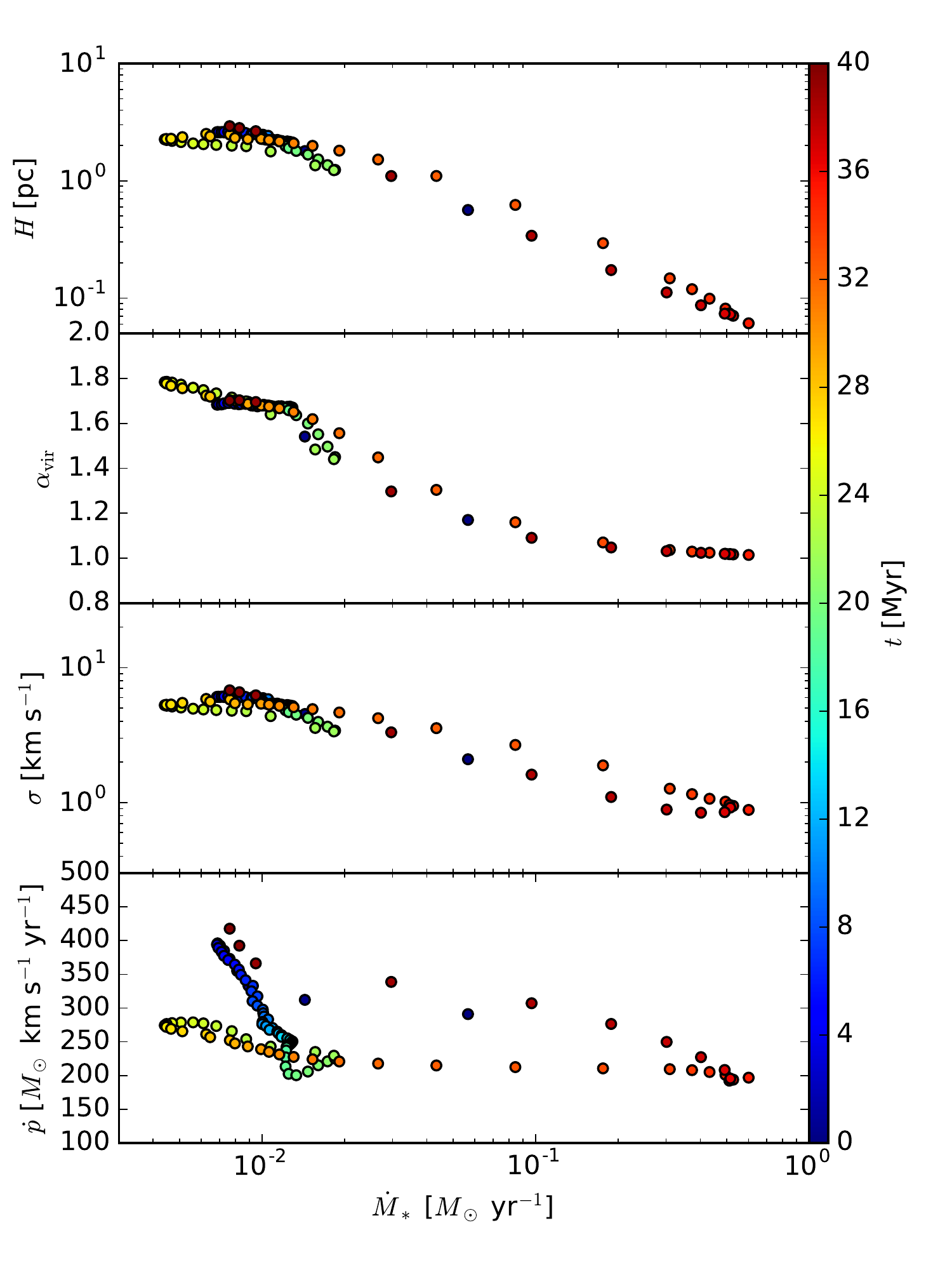}
\caption{
\label{fig:cycle_fiducial}
The cycle of star formation in the star-forming ring in run m03r050f10. The quantity shown on the horizontal axis is the total star formation rate within the star-forming region at $r = 100\pm 10$ pc. The quantities plotted on the vertical axes are the scale height $H$, virial ratio $\alpha_{\rm vir}$, velocity dispersion $\sigma$, and total momentum injection rate $\dot{p}$ in this region. We compute the first three of these quantities as averages over the ring, weighted by the star formation rate in each annulus; total momentum injection rate $\dot{p}$ and the star formation rate $\dot{M}_*$ are integrated over the ring. Each circle represents a snapshot in time separated by $0.4$ Myr, and the points plotted cover a time interval from 460 - 500 Myr of evolution. Points are coloured by time offset from 460 Myr.
}
\end{figure}

In this dead zone, gas accumulates and, as this happens, the velocity dispersion, scale height, and virial ratio all drop. Immediately outside the dead zone the scale height remains fairly constant at tens of pc and the velocity dispersion at tens of km s$^{-1}$, but inside the dead zone, the velocity dispersion drops as low as $\sim 1$ km s$^{-1}$ and the scale height reaches $\sim 1$ pc. This first occurs at $\sim 15-20$ Myr of evolution and, at this point, star formation begins. Momentum injection from star formation begins in earnest a few Myr later, and this in turn drives a wind with a mass flux comparable to the star formation rate, while also pumping up the turbulent velocity dispersion, scale height, and virial ratio, all of which lower the star formation rate. By $\sim 100$ Myr of evolution, the system has settled into a quasi-steady cycle, which we illustrate further in \autoref{fig:cycle_fiducial}. Star formation thereafter proceeds in bursts, always centred on a ring located at the shear minimum. To be quantitative, the time-averaged star formation rate peaks at $r_{\rm peak}=100$ pc, whereas the minimum of shear is at $r=81$ pc. Averaged over time, material at $r = 100 \pm 10$ pc accounts for 35\% of the mass and 48\% of the star formation in the computational domain. The velocity dispersion, virial ratio, and scale height in this region undergo cycles of increase and decay, oscillating between $\sigma \approx 1-10$ km s$^{-1}$, $H \approx 0.1 - 10$ pc, $\alpha_{\rm vir} \approx 1 - 2$. These in turn drive corresponding cycles in the depletion time, star formation rate, and momentum injection rate. 

Examining the final panel in \autoref{fig:cycle_fiducial}, one can see a clear phase shift between momentum injection and star formation: at the start of the time interval shown (blue points), the momentum injection rate is high and the star formation rate is low. After $\sim 20$ Myr the momentum injection rate declines, and after $\sim 30$ Myr the star formation rate rises while the momentum injection rate remains low. Finally, at $\sim 40$ Myr, the momentum injection rate rises again, returning to a value similar to that at the start of the cycle.

\begin{figure}
\includegraphics[width=\columnwidth]{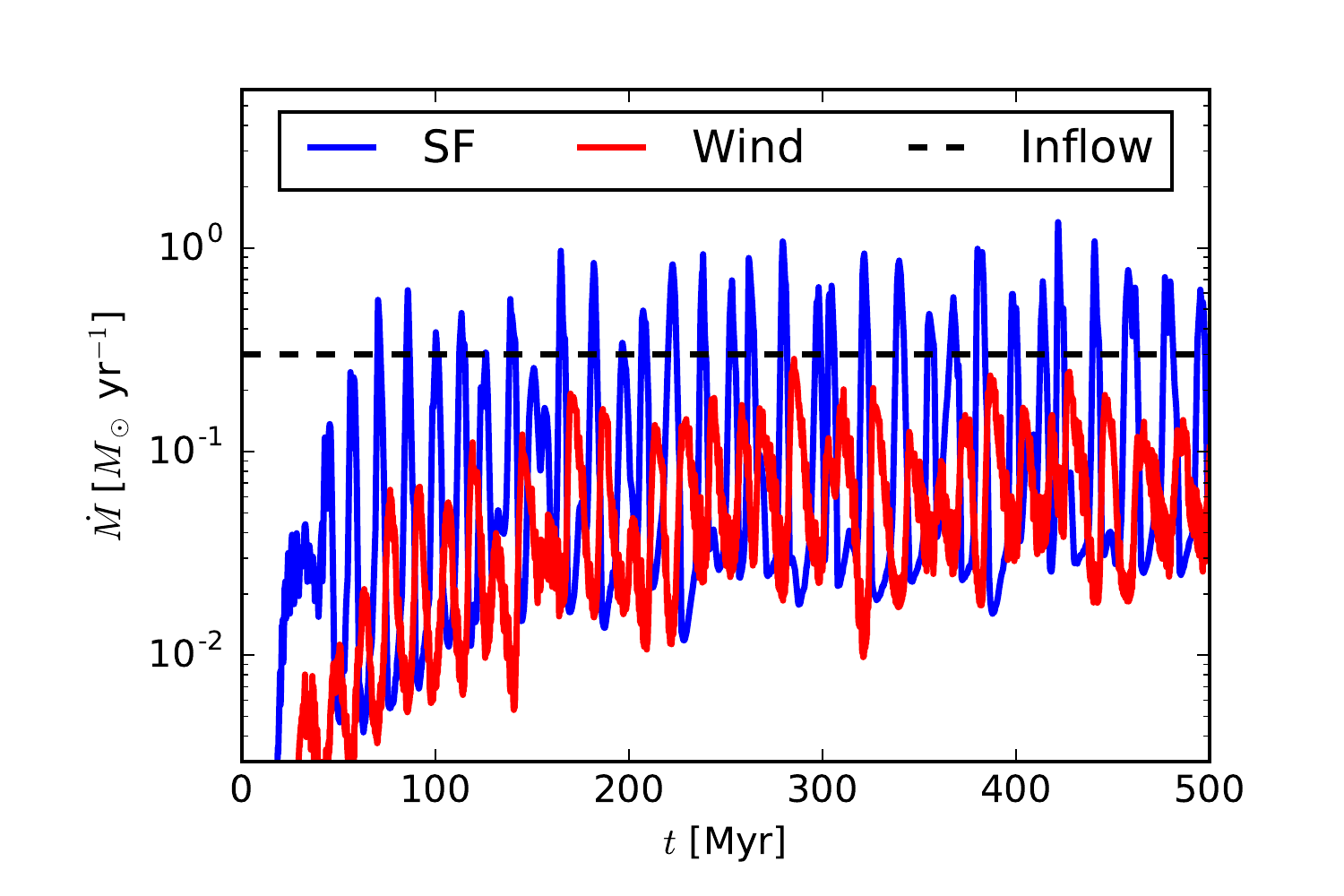}
\caption{
\label{fig:sfr_fiducial}
Area-integrated rates of star formation (solid blue line), mass loss via winds (solid red line), and mass inflow from the outer boundary (dashed black line) in run m03r050f10.
}
\end{figure}

We can see the bursts more clearly by integrating over the entire disc. In \autoref{fig:sfr_fiducial} we show the integrated rates of star formation, mass inflow, and mass outflow (via the wind) in the entire disc. it is clear that, once the system reaches quasi-equilibrium, star formation is an episodic phenomenon with a rough period of tens of Myr. The wind mass loss rate is also periodic, but with smaller oscillations than the star formation rate. Wind mass launching has a slight phase lag relative to star formation, as one might expect: winds are launched a few Myr after a peak in the star formation rate, since this is when supernova momentum injection peaks.

Before proceeding further, it is useful to distinguish between the true star formation rate, and what an observer would infer using a star formation tracer. The most common tracers for the Galactic CMZ are based on ionising photon production, and we therefore compute the total ionising luminosity produced in our simulations via
\begin{equation}
Q = \int_0^\infty{\dot{M}_*}(t-t') q(t')\, dt',
\end{equation}
where $q(t)$ is the ionising luminosity per unit mass for a simple stellar population of age $t$; we derive this quantity from the same \texttt{starburst99} computations described in \autoref{ssec:feedback}. We then convert this to a star formation rate via
\begin{equation}
\dot{M}_{*,\rm obs} = \frac{Q}{1.57\times 10^{53}\,(\mathrm{photons}\;\mathrm{s}^{-1})/(M_\odot\;\mathrm{yr}^{-1})},
\end{equation}
where the conversion factor is derived from a \texttt{starburst99} computation for a population with a constant star formation rate at an age of 50 Myr. Because $\dot{M}_{*,\rm obs}$ is derived from an integral over the stellar population, it slightly lags and smooths the true star formation rate.

\begin{figure}
\includegraphics[width=\columnwidth]{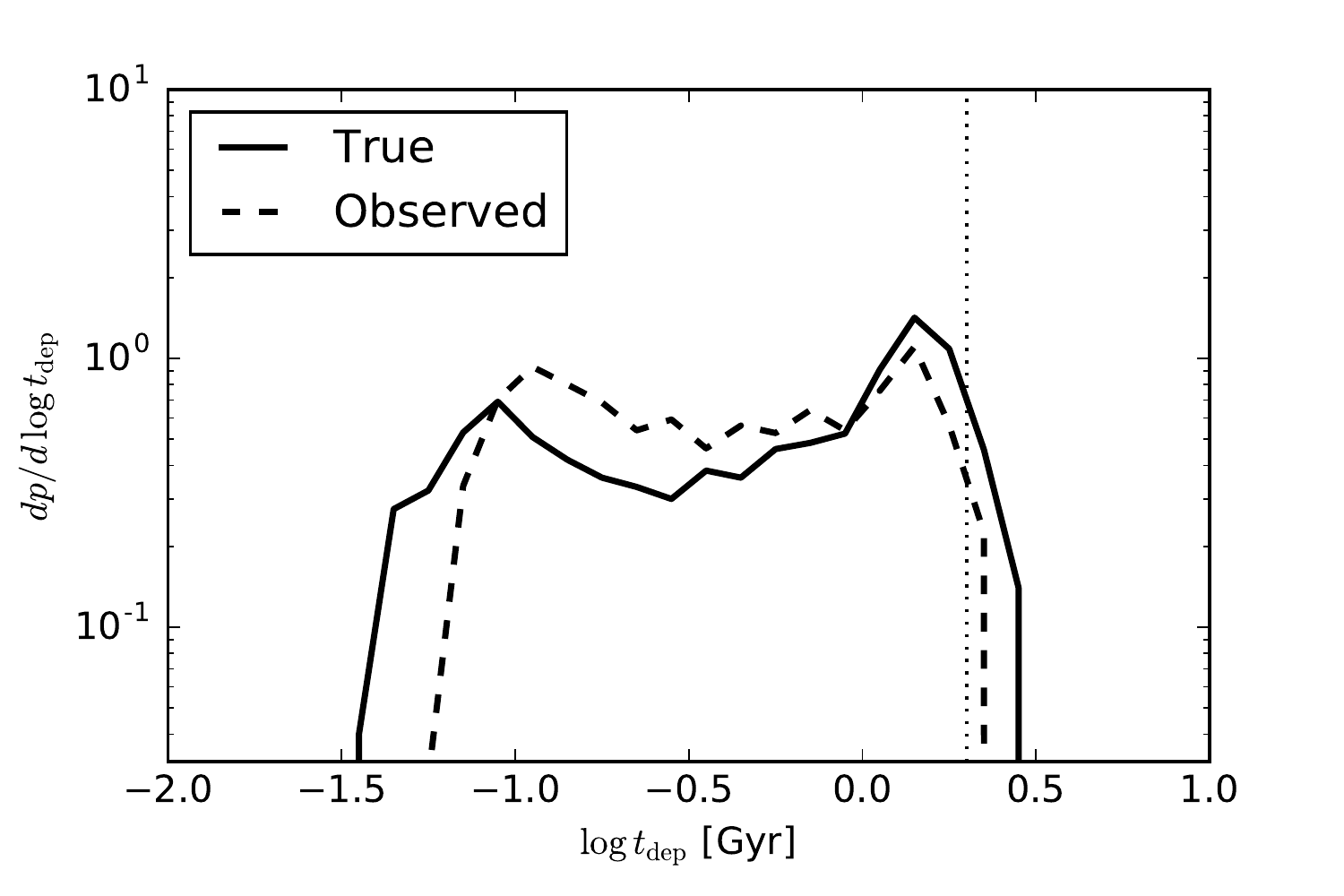}
\caption{
\label{fig:tdep_pdf_fiducial}
Probability distribution function $dp/d\log t_{\rm dep}$ for the logarithm of the instantaneous depletion time $t_{\rm dep}$, computed using both the true (solid line) and observed (dashed line) area-integrated star formation rates in run m03r10f10. The dotted vertical line shows a depletion time of $2$ Gyr.
}
\end{figure}

With this quantity in hand, in \autoref{fig:tdep_pdf_fiducial} we plot the corresponding probability distribution function (PDF) for the true and observationally-inferred log depletion times, $dp/d\log t_{\rm dep}$. This quantity is simply the probability density that the system would show a log depletion time between $\log t_{\rm dep}$ and $\log t_{\rm dep} + d\log t_{\rm dep}$ if it were observed at a random time; in computing this statistic we only consider times $t>200$ Myr, to exclude the phase when the system is still settling into steady state. We see that the system spends $\approx 40\%$ of its time with a depletion time $>1$ Gyr, a typical value for outer discs, and the other $\approx 60\%$ with a substantially smaller depletion time, which we might characterise as an outburst state. If we use the observationally-inferred star formation rate instead, these figures shift to 30\% in ``normal" star-forming mode and 70\% in ``starburst" mode; the difference arises because the use of the ionisation-based star formation rate smears out the bursts, making them appear to last longer.

\begin{figure}
\includegraphics[width=\columnwidth]{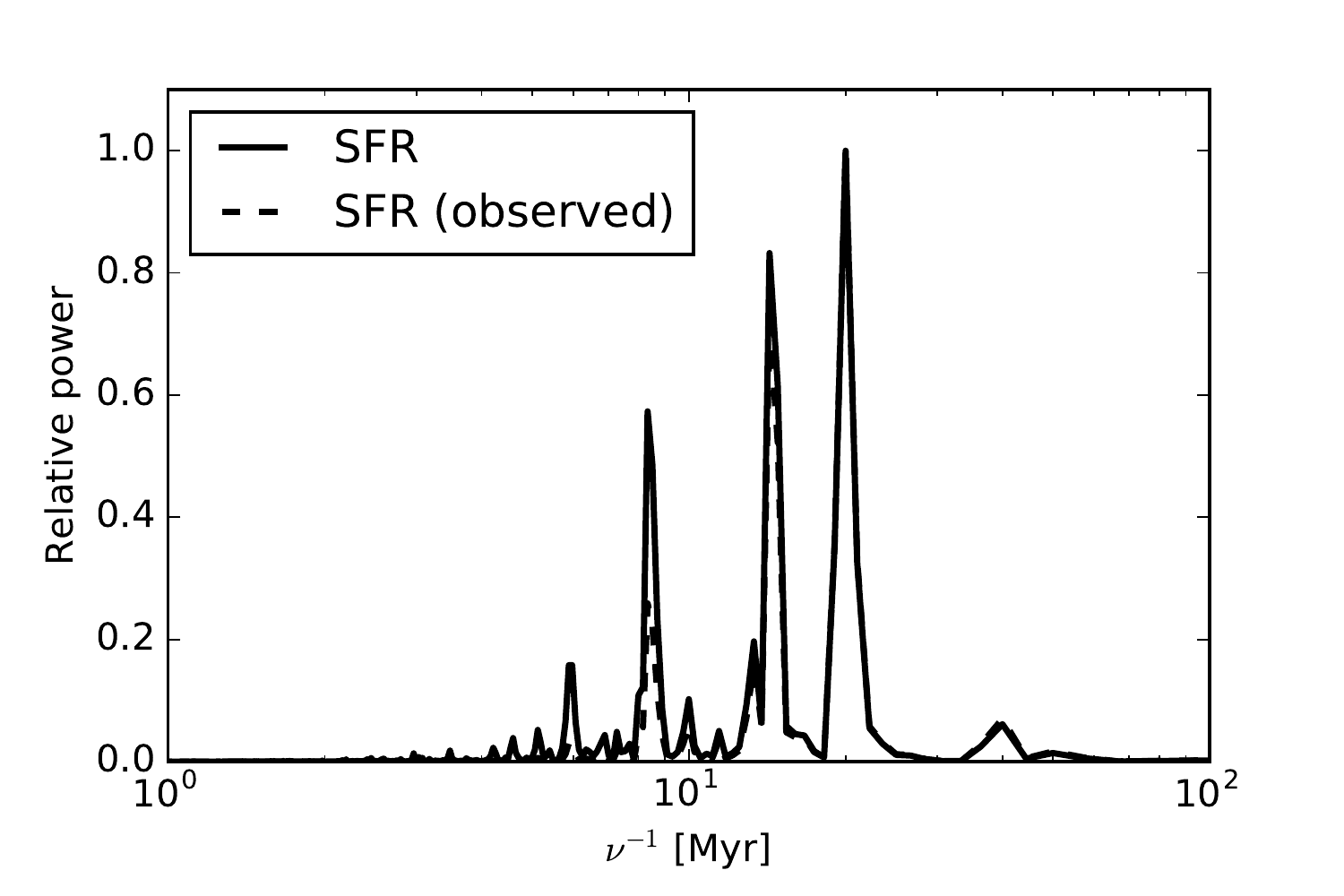}
\caption{
\label{fig:periodogram_fiducial}
Periodogram of the true (solid line) and observed (dashed line, almost completely hidden by the solid line) area-integrated star formation rates in run m01r050f10. The $x$ axis shows the period, and the $y$ axis shows power normalised by the power in the highest power bin. We compute this periodogram using a Hann window function.
}
\end{figure}

To further characterise the burst behaviour, in \autoref{fig:periodogram_fiducial} we show a periodogram of the true and observed star formation rates; modulo issues of numerical aliasing due to the finite number of samples and the non-periodicity of the data, this quantity is simply the power spectrum of the star formation history, plotted as a function of inverse frequency. From the periodogram, see that there are primary power spikes at tens of Myr, with secondary spikes at $\sim 5-10$ Myr. To make this quantitative, we define two timescales, $\nu^{-1}_{\rm min}$, and $\nu^{-1}_{\rm max}$, as the minimum and maximum inverse frequencies for which the power spectral density density $P(\nu)$ is equal to 10\% of its peak value. The choice of 10\% is somewhat arbitrary, but results are not very sensitive to the exact threshold we choose, and visual examination of the periodograms and time series shows that this choice does a good job of reproducing what one would pick out by eye. Intuitively, we may think of $\nu^{-1}_{\rm min}$ and $\nu^{-1}_{\rm max}$ as characterising the shortest and longest timescale on which the star formation rate varies, with the former describing the short duration of individual bursts, and the latter describing the longer periodicity between bursts. For run m03r050f10, we find $\nu^{-1}_{\rm min} = 5$ Myr and $\nu^{-1}_{\rm max} = 21$ Myr; if we consider the observationally-inferred star formation rate instead, the longest period remains roughly the same, while the shortest one increases to about 8 Myr.

\begin{figure}
\includegraphics[width=\columnwidth]{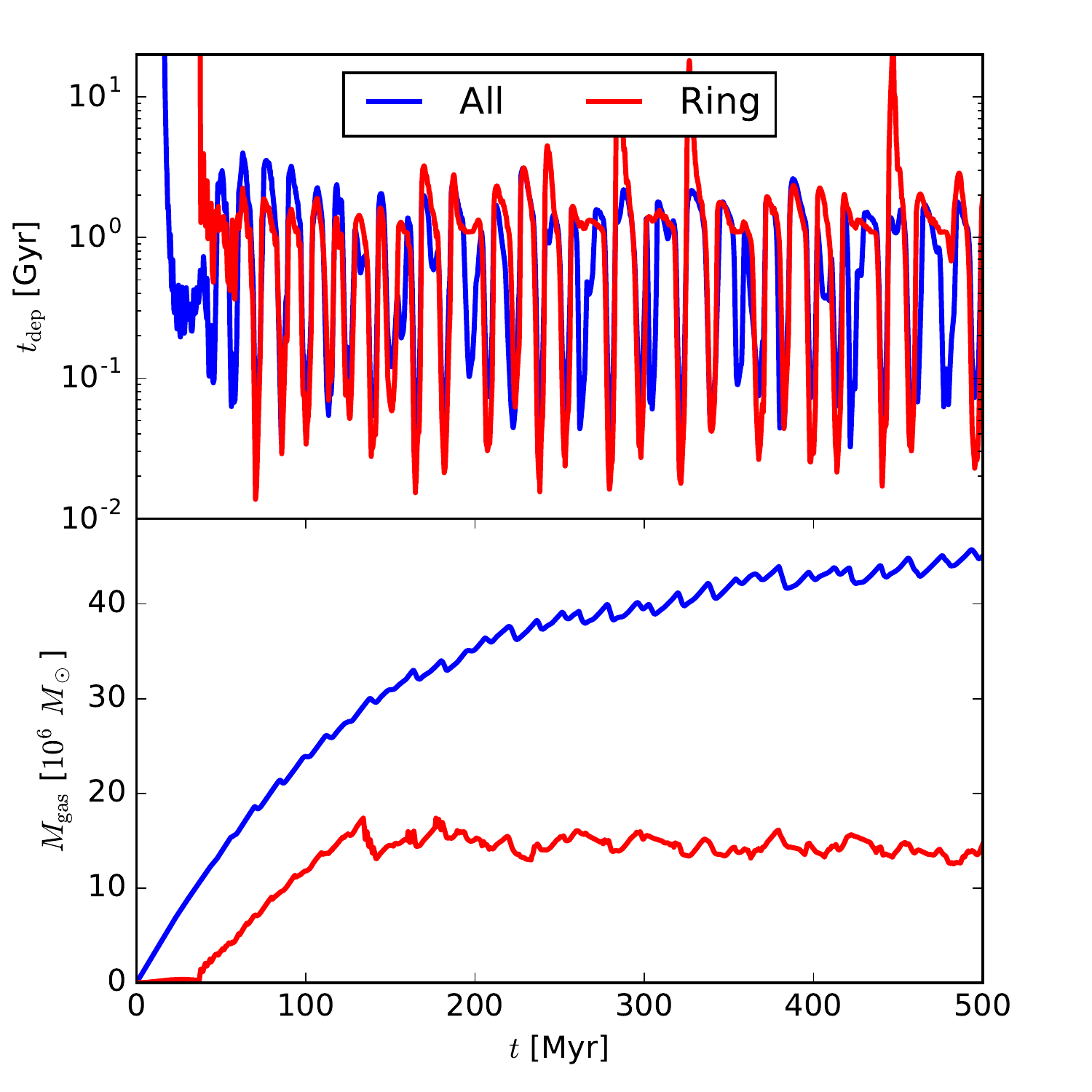}
\caption{
\label{fig:tdep_mgas_fiducial}
Depletion time (top panel) and gas mass (bottom panel) in run m03r050f10. We show results both for the entire computational domain (blue line) and for the ring of material at $r_{\rm peak}\pm 10$ pc, where $r_{\rm peak}=100$ pc is the radius at which the time-averaged star formation rate reaches its maximum.
}
\end{figure}

During an outburst the star formation rate rises by a factor of $\sim 100$, from a few percent of the gas inflow rate to several times the inflow rate. As noted above, the wind mass loss rate varies less than the star formation rate, showing only factor of $\sim 10$ changes from peak to trough. Consequently, there is wide variation in the ratio of the wind outflow rate to the star formation rate, known as the mass loading factor. This can be as high as $\sim 10$ immediately after an outburst, and as small as $\sim 0.3$ immediately after a burst begins, before supernovae begin to occur. Averaging over all times after $200$ Myr, once the gas mass in the system reaches steady state, we find that the star formation efficiency is
\begin{equation}
\label{eq:SFE}
\mathrm{SFE} = \frac{\left\langle\dot{M}_*\right\rangle}{\left\langle\dot{M}_* + \dot{M}_{\rm wind}\right\rangle} = 0.72,
\end{equation}
where the angle brackets indicate an average over times $>200$ Myr. Note that, since the total gas mass in the CMZ in our model is in steady state, the star formation efficiency is simply the mean fraction of the gas that enters the CMZ that is ultimately converted to stars. In our fiducial model, we find that 72\% of the inflowing mass goes into stars, while the remaining 28\% is lost in the form of winds. Phrased in terms of a mass loading factor, this corresponds to a time-averaged mass loading factor of 0.39.

What causes the bursts? As limiting cases we could imagine that changes in the star formation rate are driven by changes in the gas mass while the gas depletion time remains fairly constant, changes in the depletion time while the gas mass remains fairly constant, or some combination of the two. To address this question, in \autoref{fig:tdep_mgas_fiducial} we show the total gas mass and the gas depletion time measured from the simulations. We show these quantities both for the entire computational domain, and for the ring of material at $r_{\rm peak}\pm 10$ pc. The Figure clearly shows that, while there is some periodic variation in the gas mass, it is far smaller than the variation in the depletion time. Thus bursts are not caused by wholesale ejection of mass from the ring, though \autoref{fig:summary_fiducial} shows that there clearly are local evacuations. Instead, they are caused when the gas is driven to higher velocity dispersions by the effects of stellar feedback. This in turn lowers the star formation rate, both by raising the virial ratio and by increasing the gas scale height and thus lowering the density. After several tens of Myr, the momentum injection rate drops and is no longer able to sustain the high level of turbulence. The velocity dispersion decreases and another outburst cycle begins. We note that this form of the feedback cycle is contrary to what we proposed in \citetalias{krumholz15d}, where we conjectured that there would be wholesale ejection.

\subsection{Effects of Varying $\epsilon_{\rm ff,0}$}

\begin{figure*}
\includegraphics[width=\textwidth]{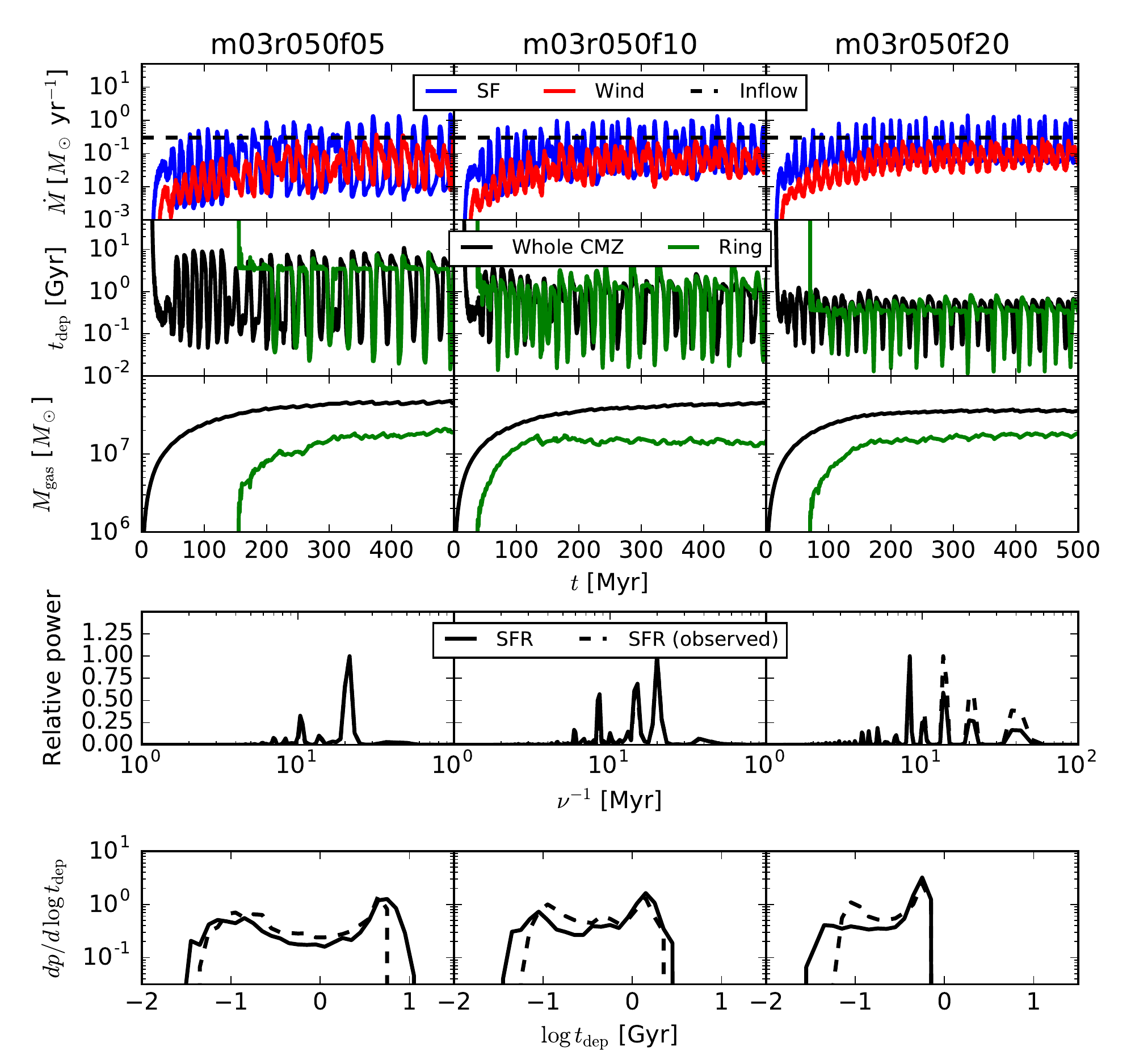}
\caption{
\label{fig:compare_epsff}
Results from runs m03r050f05, m03r050f10, and m03r050f20 (left, centre, and right columns), which all have the same gas inflow rate and feedback prescription, but differ in the parameter $\epsilon_{\rm ff,0}$ that describes the star formation rate per dynamical time. The top three rows show, as a function of time, from top to bottom: mass per unit time $\dot{M}$ converted to stars (blue), lost in winds (red), and entering the outer edge of the disc (black dashed); gas depletion time $t_{\rm dep}$ computed over the entire disc (black) and over the ring of peak star formation at $r = 100\pm 10$ pc (gren); gas mass $M_{\rm gas}$ in the entire disc (black) and the star-forming ring (green). The bottom two rows show the periodogram of the star formation rate and the PDF of depletion times, for both the true (solid black) and observed (dashed black) star formation rates. All quantities are computed in exactly the same manner as those shown in \autoref{fig:sfr_fiducial} - \autoref{fig:tdep_mgas_fiducial}, and the data in the central column here are identical the data shown in those figures.
}
\end{figure*}

The parameter $\epsilon_{\rm ff,0}$, which controls the star formation rate per free-fall time, is significantly constrained by observations. However, there are uncertainties nonetheless, not the least because the virial ratios of observed objects are generally only determinable to the factor of $\sim 2-3$ level. For this reason we compare runs m03r050f05, m03r050f10, and m03r050f20, in which we fix all parameters but $\epsilon_{\rm ff,0}$, which we vary from $0.005 - 0.02$. We compare the star formation histories, depletion times, gas masses, periodograms, and depletion time PDFs in \autoref{fig:compare_epsff}.

Examining the different columns in the Figure, we see that all runs again show very similar qualitative behaviour. The main effect of varying $\epsilon_{\rm ff,0}$ is to change the amplitude and timescale of the periodic variation. Using $\epsilon_{\rm ff,0}=0.005$, a factor of two smaller than our fiducial case, leads to a star formation rate and wind outflow rate that varies somewhat more slowly but with somewhat larger amplitude than our fiducial case, while $\epsilon_{\rm ff,0}=0.02$ produces more frequent oscillations of smaller magnitude. This is reflected in the spread in the depletion time PDF as well, with higher values of $\epsilon_{\rm ff,0}$ leading to shorter depletion times especially during quiescence, thereby compressing the overall range of depletion times achieved. With $\epsilon_{\rm ff,0} = 0.02$, the depletion time never rises above 1 Gyr, which is incompatible with observation of galactic centres.

However, the time variation of the star formation rate, as measured by the periodogram, is very similar in the three runs, as is the time-averaged star formation efficiency. The exact values of $\nu_{\rm min}^{-1}$, $\nu_{\rm max}^{-1}$, and SFE are given in \autoref{tab:sims}. Clearly, the periodic behaviour we observe is insensitive to the exact value of $\epsilon_{\rm ff,0}$.

\subsection{Effects of Varying $\epsilon_r$}

The most uncertain value in our model is $\epsilon_r$, the radius over which stellar feedback is spread due to the fact that newly-formed stars are not on orbits that are identical to those of the gas. We have argued that it should be approximately $0.05$ based on computing the spreads in orbits we expect based on the observed velocity dispersions of CMZ star clusters, but these velocity dispersions are purely empirical inputs, and could conceivably have been different at different times or in different galaxies. To test how uncertainty in $\epsilon_r$ might affect our conclusions, in runs m03r025f10, m03r050f10, and m03r100f10, we hold all parameters except $\epsilon_r$ fixed, and vary the value of $\epsilon_r$ from $0.025$ to $0.1$ (see \autoref{tab:sims}).

\begin{figure*}
\includegraphics[width=\textwidth]{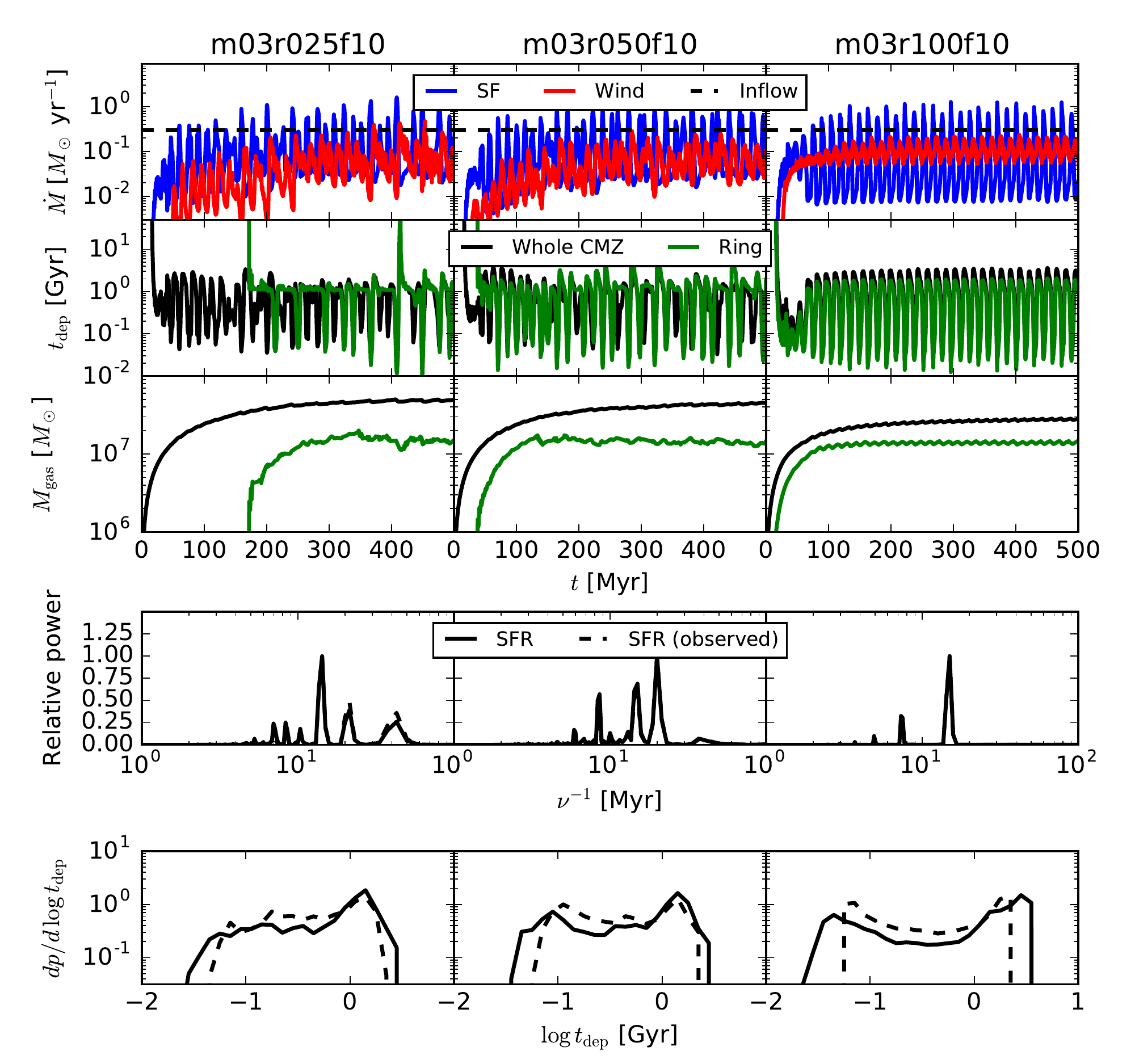}
\caption{
\label{fig:compare_er}
Same as \autoref{fig:compare_epsff}, but now comparing runs m03r05f10, m03r10f10, and m03r20f10, which use identical values for the parameters describing star formation and inflow, but differ in the radial extent over which feedback is injected.
}
\end{figure*}

We show the results of the three simulations with varying $\epsilon_r$ in \autoref{fig:compare_er}. Qualitatively, runs m03r025f10 and m03r050f10, corresponding to $\epsilon_r = 0.025$ and $0.05$, are nearly identical. Run m03r100f10, corresponding to $\epsilon_r = 0.1$, also shows bursty behaviour, but its periodicity is much more regular than in the other two runs. While the PDF of depletion times is much the same, the periodogram for this run shows a single dominant peak at about 20 Myr, rather than several peaks as shown in our other runs. Examining the spatial distribution of gas and star formation for this run (not shown), we see that, rather than a patchy and irregular pattern of star formation found in the other runs, it has a more regular morphology, with star formation always occurring at the same radial location. Nonetheless, we find that burstiness on $\sim 20$ Myr timescales is again a generic outcome of the simulations.

We can understand these results, and in particular the difference between the $\epsilon_r = 0.1$ and smaller $\epsilon_r$ cases, by thinking about the spread in feedback compared to the width of the star-forming ring. As noted above, in our fiducial case close to 50\% of the star formation takes place within a ring of $\pm 10$ pc width about $r=100$ pc, so the fractional width of the star-forming region is roughly 10\%. For $\epsilon_r < 0.1$, the feedback is localised within the star-forming ring, and causes disruption of patches of it. This leads to the chaotic, bursty behaviour we observe for the $\epsilon_r = 0.025$ and $0.05$ cases. On the other hand, for $\epsilon_r \gtrsim 0.1$, the feedback becomes close to uniform across the star-forming ring. This reduces its effectiveness somewhat, since some of the momentum is delivered to the non-star-forming gas outside the ring, and also means that the star-forming region reacts coherently rather than locally to the feedback. This coherent response explains the regular pattern we observe in star formation rates for the $\epsilon_r = 0.1$ case.

\subsection{Effects of Varying $\dot{M}_{\rm in}$}

\begin{figure*}
\includegraphics[width=\textwidth]{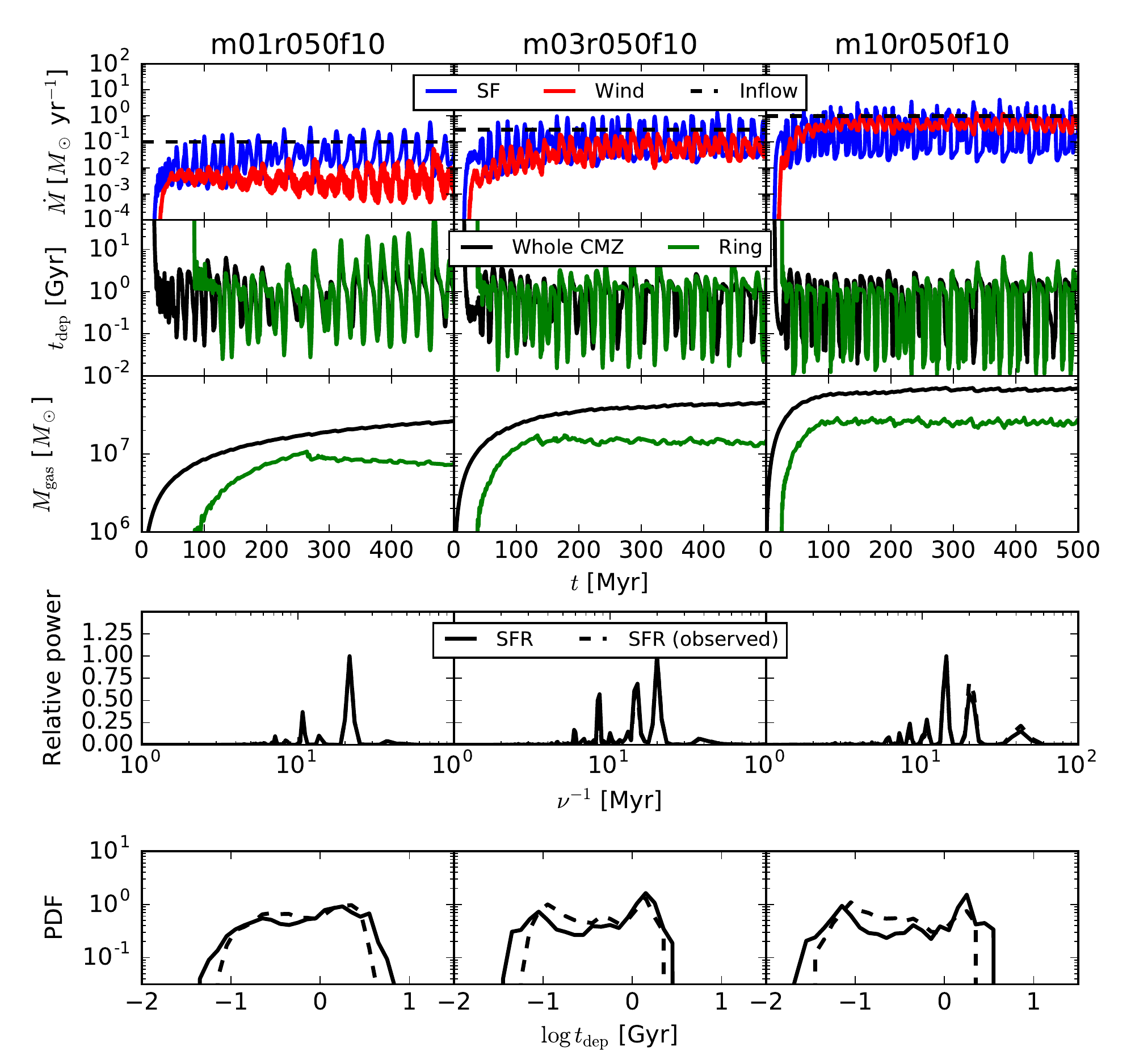}
\caption{
\label{fig:compare_mdot}
Results from runs m01r050f10, m03r050f10, and m10r050f10 (left, centre, and right columns). These all have the same parameters for star formation and feedback, but differ in the mass inflow rate into the CMZ that we assume, with values of $\dot{M}_{\rm in} = 0.1$, $0.3$, and $1.0$ $M_\odot$ yr$^{-1}$, respectively. Panels are the same as in \autoref{fig:compare_er}.
}
\end{figure*}

The final parameter of our model that we consider varying is $\dot{M}_{\rm in}$, the mass accretion rate onto the CMZ from outside. We explore how this parameter affects the behaviour of the CMZ via models m01r050f10, m03r050f10, and m10r050f10, where we use values of $\dot{M}_{\rm in}=0.1$, 0.3, and $1.0$ $M_\odot$ yr$^{-1}$, while holding all other parameters fixed. We show the results of this experiment in \autoref{fig:compare_mdot}.

Examining the Figure, it is clear that the primary quantities that are influenced by the inflow rate are, not surprisingly, the star formation and wind mass ejection rates, and steady-state gas mass of the CMZ and in the 10 pc ring. All these quantities appear to scale nearly linearly with the inflow rate. The temporal pattern of star formation is qualitatively the same in all the runs. The main systematic difference we see is in the partition of the inflow between star formation and winds. At the lowest inflow rate, $0.1$ $M_\odot$ yr$^{-1}$, the star formation rate exceeds the mass outflow rate at essentially all times, leading to a comparatively high star formation efficiency of $\approx 90\%$. In contrast, at an inflow rate of $1.0$ $M_\odot$ yr$^{-1}$ the wind outflow rate and star formation rate are nearly the same, leading to a star formation efficiency closer to 50\%.

\section{Discussion}
\label{sec:discussion}

\subsection{A Dynamical Model of the CMZ}

We are now in a position to make some general statements about how star formation in the Milky Way's CMZ, and the analogous regions of other galaxies, should behave. Gas enters the CMZ as a result of transfer by the Galactic Bar, and the bar further drives instabilities in the region of high shear that transport mass and keep the gas too turbulent to form stars. This ends where the rotation curve switches from flat to near solid-body, and gas accumulates in this region, forming a persistent ring-like structure. Within the ring star formation occurs in bursts. The driving feature of the bursts is an alternating cycle whereby turbulence decays, leading to high densities and low virial parameters, both of which boost the star formation rate. This leads to the formation of a large stellar population, which begins producing supernovae a few Myr later, raising the level of turbulence and driving the star formation rate back down. The low rate of star formation continues for a while, but over time the supernovae fade and the turbulence decays, causing the cycle to repeat. At the same time, the supernova feedback drives a wind off the CMZ, which carries away a portion of the mass that enters. A key requirement for this cycle to take place is that the timescale for turbulent decay and the onset of star formation is shorter than the time required for the onset of supernova feedback, which prevents the system from reaching an equilibrium in which injection of energy by supernovae balances dissipation. This condition is satisfied in the CMZ, because in the low-shear region at 100 pc, the orbital period is only $\approx 3$ Myr.

Based on our simulations, we can make the following quantitative predictions about this cycle, which are robust against variations in any of our uncertain parameters.
\begin{itemize}
\item We predict that the duration of outbursts should be $\sim 5-10$ Myr, while the overall cycle of burst and quiescence should have a period of $\sim 15-40$ Myr. The former number comes mostly from the delay between the onset of star formation and the first supernovae, while the latter comes from the time required for supernovae to cease and for turbulence to decay, allowing gas to become gravitationally unstable again.
\item Throughout this cycle there is a persistent, dense gas structure at $\sim 100$ pc from the Galactic Centre, where the Galactic rotation curve begins to turn toward solid body and the shear reaches a minimum. The mass in this structure varies periodically, and local patches of it may be evacuated by feedback, but the overall variation in mass in this structure is far smaller than the variation in the star formation rate. Instead, changes in the star formation rate are driven primarily by changes in the mean density and velocity dispersion of this structure, which combine to produce a short depletion time during outbursts and a long one during quiescence.
\item During quiescence, the gas depletion time of the CMZ is of order 1 Gyr. During outburst this drops by a factor of $\sim 10-100$, reaching $\lesssim 100$ Myr. The true depletion time is $>1$ Gyr (i.e., comparable to what is seen in outer galaxies) for roughly $40\%$ of the time, and is shorter, indicating a starburst, about 60\% of the time. However, because of the short durations of the bursts, and because the true time-averaged depletion time is only a few hundred Myr, an observationally-determined fraction of the time spent in starburst will depend on the effective integration time of the star formation rate tracer used. If one measures with an ionisation-based star formation tracer, the CMZ spends $\sim 30\%$ of its time with what appears to be a ``normal" depletion time $>1$ Gyr, and $\sim 70\%$ with a shorter depletion time. The longer the integration time, the more time will appear to be spent in outburst.
\item At an inflow rate of $0.3$ $M_\odot$ yr$^{-1}$, a slight majority of the gas entering the CMZ is converted to stars, while the rest is ejected in a wind driven primarily by supernova feedback. The balance between star formation and wind loss depends mildly on the inflow rate, with lower inflow rates producing higher star formation efficiencies and higher inflow rates producing lower ones. This wind is launched primarily from the same dense structure where star formation occurs, and carries away a time-averaged mass flux that is slightly smaller than the flux of mass going into stars. However, the ratio of wind mass flux to star formation rate undergoes extreme variations, ranging from $\sim 10$ to $\sim 0.03$ depending on where the system is in the outburst cycle.
\end{itemize}

\subsection{Comparison to the Observed Milky Way CMZ}

\begin{figure}
\includegraphics[width=\columnwidth]{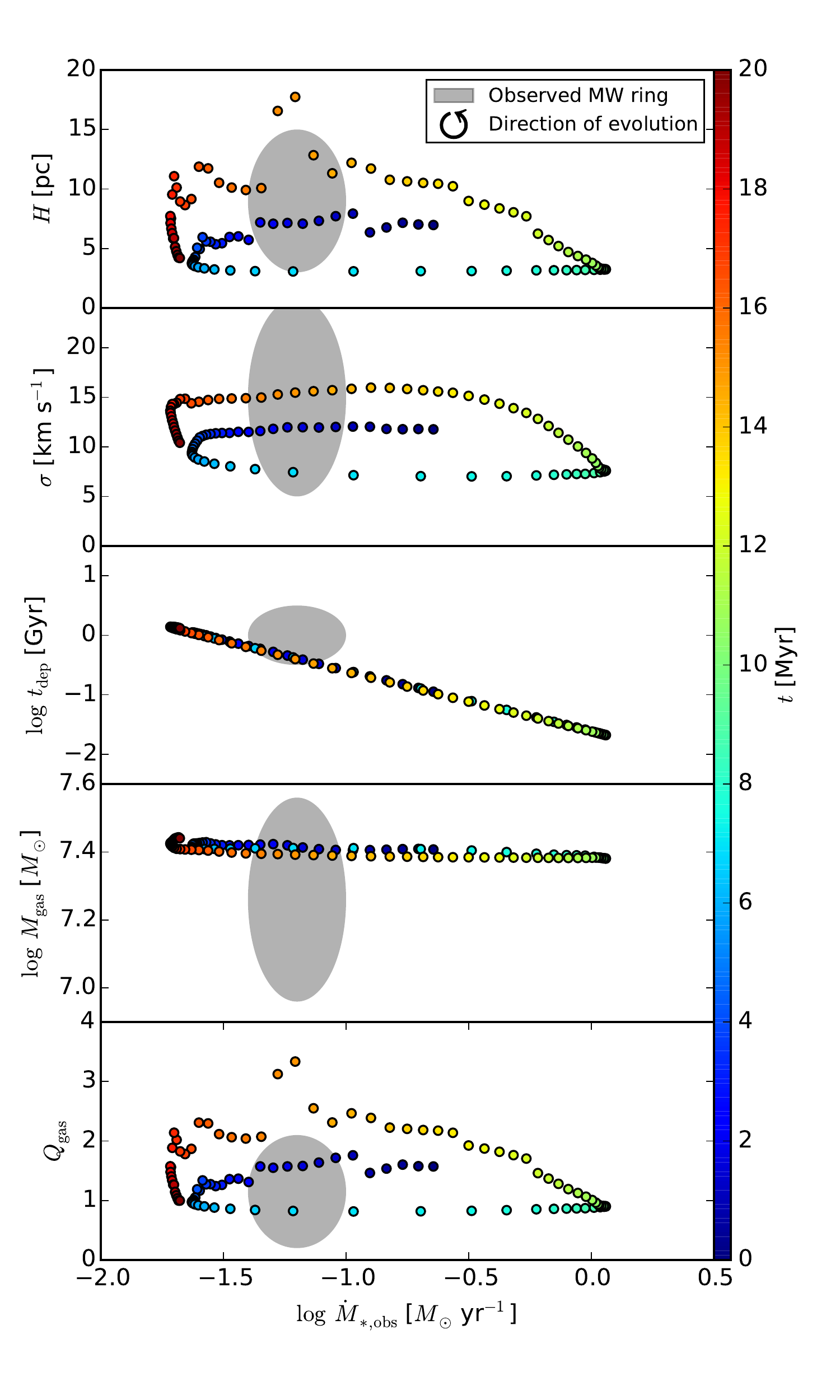}
\caption{
\label{fig:obs_cycle}
Observable cycle of properties of the star-forming ring at $100\pm 10$ pc in run m10r050f10 from $480-500$ Myr (coloured points) as compared to the observed Milky Way ring (gray ellipses). Points are coloured by time since 480 Myr, as in \autoref{fig:cycle_fiducial}, with one point per $0.2$ Myr of time. The properties shown are, from top to bottom: area-weighted mean scale height $H$, mass-weighted mean velocity dispersion $\sigma$, gas depletion time $t_{\rm dep}$, total gas mass $M_{\rm gas}$, and mass-weighted mean Toomre $Q$ parameter for the gas, where at each radius we have $Q = \kappa \sigma/\pi G \Sigma$, where $\kappa$ is the epicyclic frequency. All quantities are shown as a function of the star formation rate as observed with an ionisation-based tracer. The observational constraints shown in gray are taken from the compilations in \citet{kruijssen14b}, \citet{longmore12a,longmore13a}, and \citet{henshaw16a}; in cases where the authors did not state an uncertainty, we have adopted an uncertainty of a factor of 2.
}
\end{figure}

How do our models compare to what we actually observe in the Milky Way? To address this, we focus on model m10r050f10, which by a variety of metrics appears to be the closest match to the Milky Way's CMZ. We illustrate this in \autoref{fig:obs_cycle}, which compares various observable properties of the star-forming ring at $r=100\pm 10$ pc in the simulation to the same properties observed in the Milky Way's star-forming ring, as summarised in Table 1 of \citet{kruijssen14b} and Table 2 of \citet{longmore13a}. This figure is analogous to \autoref{fig:cycle_fiducial}, except that since we are interested in observable rather than intrinsic properties, we slightly modify the quantities plotted; for example, we use the observable rather than the true star formation rate, and we plot an area-weighted rather than a star formation-weighted scale height. \autoref{fig:obs_cycle} clearly shows that this run spends a significant amount of time with properties that closely resemble those of the observed star-forming ring in the Milky Way's CMZ.\footnote{While this discussion focuses on the star-forming ring, we note that the macroscopic properties of the models are also consistent with the observed, large-scale spatial distribution of the gas in the CMZ. \citet{longmore13a} find that a dust-inferred gas mass of $1.8\times10^7~{\rm M}_\odot$ resides within $|\ell|<1^\circ$ (or $r<140$~pc), with $2.3\times10^7~{\rm M}_\odot$ residing at $|\ell|>1^\circ$. The masses shown in \autoref{fig:obs_cycle} reproduce the observed gas mass in the inner CMZ to within the uncertainties and our predicted gas mass outside of the star-forming ring of $\sim1.4\times10^7~{\rm M}_\odot$ also provides a good match to the observed mass. Due to the high virial ratio of the gas in the outer CMZ, its scale height is predicted to be substantially larger than that of the star-forming ring ($H\sim70$~pc rather than $H\sim10$~pc). Again, this increase is qualitatively consistent with observations. At $|\ell|>1^\circ$, the total vertical extent of the observed molecular gas emission (traced by $^{12}$CO, see Figure~3 of \citealt{bally10a}) covers more than a degree in latitude (i.e.~more than 140~pc). This increase of the scale height with radius is of the same order as predicted by our models (see \autoref{fig:summary_fiducial}, as well as Figure 13 of \citetalias{krumholz15d}).} At face value, two evolutionary stages in the modelled cycle seem to match best. At a time of $t=(482.0,486.9)$ Myr, the star-forming ring in this model has $\dot{M}_{*,\rm obs} = (0.050,0.061)$~$M_\odot$ yr$^{-1}$, $H = (7.1,3.1)$~pc, $\sigma = (11.8,7.4)$~km~s$^{-1}$, $t_{\rm dep} = (0.53,0.42)$~Gyr, $M_{\rm gas} = (2.65,2.58)\times 10^7$~$M_\odot$, and Toomre $Q$ parameter $Q_{\rm gas} = (1.55,0.82)$; all of these properties match the properties of {\it some part of} the observed star-forming ring in the Milky Way within the observational uncertainties.

Both of the above two model snapshots are close to the star formation minimum in the cycle, but their evolutionary states do differ. At $t=482.0$~Myr (Case A), the star formation rate is decreasing, as the modelled star-forming ring has experienced a starburst some 5 Myr earlier (at $t=477$ Myr) and will evolve through the star formation minimum in another 3 Myr (at $t=485$ Myr), with the next star formation peak expected in 7 Myr (at $t=489$ Myr). By contrast, in the model snapshot at $t=487$ Myr (Case B), the star formation rate is rapidly increasing, as it is exactly midway between the star formation minimum at $t=485$ Myr and the maximum at $t=489$ Myr, with the most recent starburst 10 Myr earlier. If Case A best describes the star-forming ring in the Milky Way's CMZ, then the formation of the Arches and Quintuplet clusters (with ages of 3.5 and 4.8 Myr, respectively, see \citealt{schneider14a}) has taken place at the height of the most recent starburst. However, the highest-density clouds in the star-forming ring (all situated on the `dust ridge' between Sgr A$^*$ and Sgr B2, which has enough mass to form several Arches-like clusters, \citealt{longmore13b}) have such low velocity dispersions ($<10$~km~s$^{-1}$) and small scaleheights (few pc) that they best match the conditions of Case B \citep[cf.][]{henshaw16a}. In other words, our model predicts that these clouds are unlikely to remain quiescent for another 7 Myr (as would be required in Case A). These points suggest that the star-forming ring in the Milky Way's CMZ has a non-zero spread in evolutionary times in the cycle of \autoref{fig:obs_cycle}. This is not surprising; gas is continuously spiralling onto the star-forming ring, implying that a natural time interval for an evolutionary spread is the orbital time of the gas streams within the ring. The time-scale is $\sim4$~Myr \citep{kruijssen15a} and provides a good match to the time difference between these two best-fitting model snapshots. This does, however, point to a limitation of our axisymmetric assumption.

This scenario has the following implications:
\begin{enumerate}
\item
The star-forming ring covers the entire timeline between $t=482$--$487$~Myr and {\it on average} resides at the star formation minimum ($t=485$ Myr).
\item
The previous starburst took place at $t=477$ Myr, some $\sim8$ Myr ago.
\item
The Arches and Quintuplet clusters represent the last clusters that formed during this previous starburst ($\sim5$~Myr ago).
\item
The dust ridge contains the clouds that will collapse and form stars first during the onset of the upcoming starburst (in 1--2 Myr).
\item
The non star-forming, high-velocity dispersion, and large-scale height gas in the star-forming ring has recently been accreted.
\end{enumerate}

\begin{figure}
\includegraphics[width=\columnwidth]{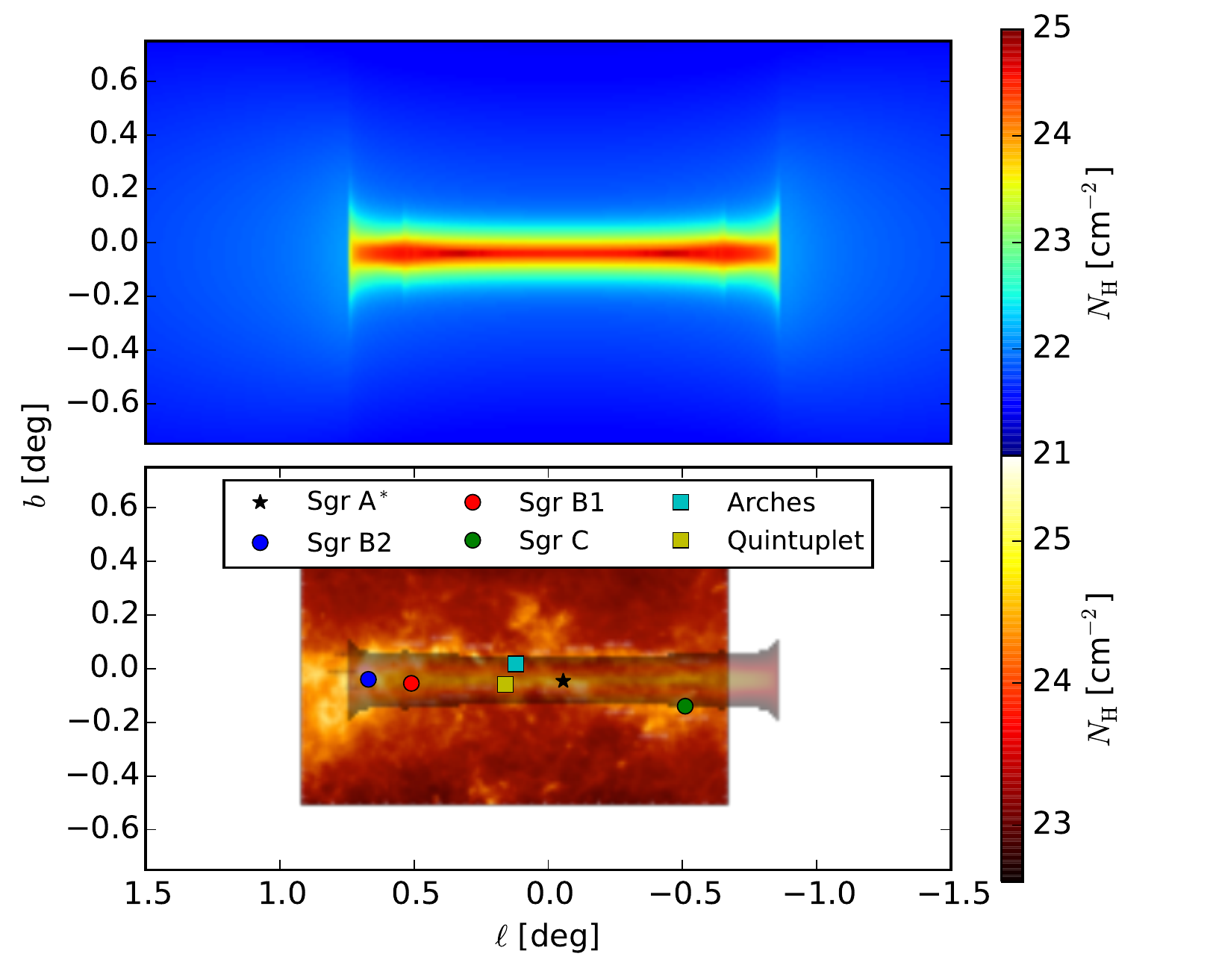}
\caption{
\label{fig:galmap}
Top panel: column density $N_{\rm H}$ for model m10r050f10 at time $t=485$ Myr as seen from Earth, with positions indicated in Galactic longitude $\ell$ and Galactic latitude $b$. Bottom panel: the image shows the column density map of \citet{molinari11a}, derived from \textit{Herschel} observations. We have superimposed on it the column density map shown in the top panel, with the colour scale adjusted to match that used in the \textit{Herschel} map; only pixels with $N_{\rm H}>4\times 10^{22}$, the minimum column in the \textit{Herschel} map, are shown, and those that we do show have been left partially transparent to allow comparison with the underlying image. The coloured circles in the lower panel mark the locations of the star-forming molecular clouds Sgr B2, Sgr B1, and Sgr C, as indicated; coordinates for these structures are taken from Table 3 of \citet{henshaw16a}. The coloured squares mark the positions of the Arches and Quintuplet clusters, and the black star indicates the position of Sgr A$^*$.
}
\end{figure}

\begin{figure}
\includegraphics[width=\columnwidth]{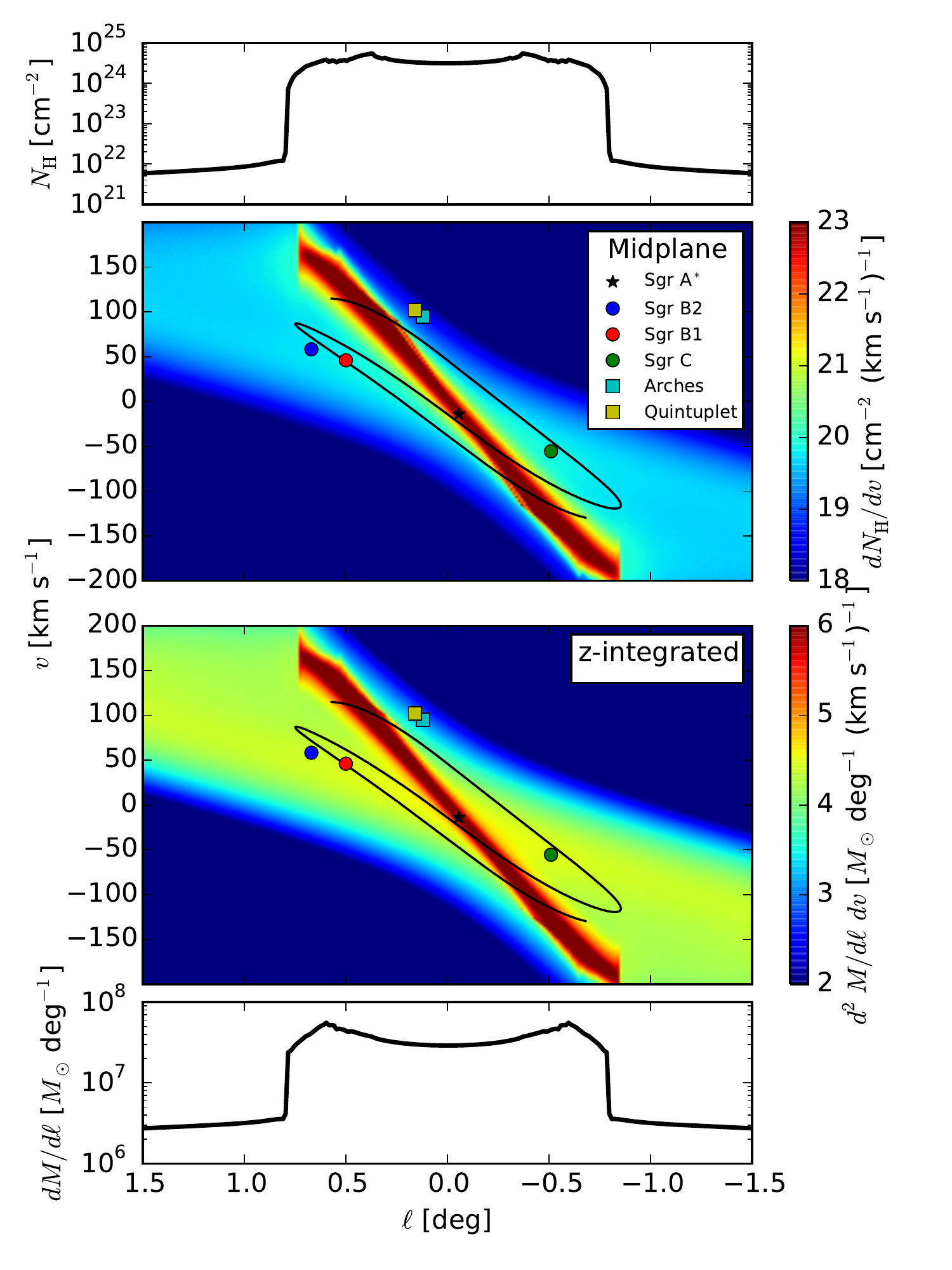}
\caption{
\label{fig:pv_plot}
Position-velocity diagram for the same snapshot as shown in \autoref{fig:galmap}. The top colour panel, labeled ``Midplane", shows the column density per unit velocity along a cut through the Galactic midplane. The line plot above this shows the velocity-integrated column density corresponding to this. The bottom coloured panel, labeled ``z-integrated", shows the total mass per unit velocity per degree of Galactic longitude, integrating over Galactic latitude. The line plot below it shows the mass per degree of Galactic longitude, integrating over the velocity and Galactic latitude. The coloured circles and squares in the coloured panels indicate the positions and velocities of Sgr A$^*$, Sgr B2, Sgr B1, Sgr C, Arches, and Quintuplet, as indicated. The black line in the coloured panels shows the best-fitting orbital model describing the observed kinematics of the gas stream from \citet{kruijssen15a}, which highlights the difference in kinematics expected when the orbit is eccentric (with $e=0.3$). Data are from the same sources as in \autoref{fig:galmap}.
}
\end{figure}

We further analyse the observable properties of the snapshot at $t=485$~Myr (corresponding to the average evolutionary phase of the star-forming ring, i.e., at a star formation minimum). To do so, we generate synthetic column density and position-velocity maps for the model, placing them in Galactic coordinates for the purposes of comparing to the observed CMZ. For the purposes of this calculation we assume a Galactic Centre distance of $8.5$ kpc \citep{ghez08a}, and following \citet{kruijssen15a} and \citet{henshaw16a} we place the center of our simulated disk at the position of Sgr A$^*$, which in position-position-velocity space is $(\ell, b, v) = (-0.\!\!^\circ056, -0.\!\!^\circ047, -14.0\mbox{ km s}^{-1})$.

We show the column density map for our model, overlaid with the observed column density distribution in the CMZ from \citet{molinari11a}, in \autoref{fig:galmap}. The figure displays an impressive degree of agreement. The predicted vertical gas distribution quantitatively matches the observations, and the extent of the ring is nearly correct as well. The locations of the most active sites of ongoing star formation -- Sgr B2, Sgr B1, and Sgr C -- are exactly where the model predicts that such sites should be found.

We show the synthetic position-velocity diagram in \autoref{fig:pv_plot}. This can be compared to observations such as those presented by \citet[their Figures 6-9]{henshaw16a}. In comparison, we find that the positional extent of our model is in good agreement with the data, as one would have expected based on \autoref{fig:galmap}, but the model velocities are somewhat higher than the observed ones. This is also apparent in the offset between the velocities of the Sgr B2, B1, and C molecular clouds and our data.

The discrepancy between model and data reflects the limitations of our assumption of axisymmetry, which requires only circular orbits. In reality, \citet{kruijssen15a} show that the star-forming ring is only partially filled with dense gas, which orbits in an open stream with non-zero eccentricity $e\approx 0.3$. The extent of this orbit is from $\approx 60-120$ pc, in excellent agreement with our model, but in the unstable region we do not have filled circular orbits, but instead partially-filled elliptical ones. The orientation is such that the Sgr molecular clouds, and the bulk of the dense gas, lie near the apocenters of the orbit, producing line of sight velocities substantially smaller than the circular velocities at their projected positions. In addition, the eccentric nature of the orbit leads to orbital precession, which manifests itself as a vertical drift in the position-velocity space of \autoref{fig:pv_plot}. A comparison to the orbital model of \citet{kruijssen15a} shows that the Sgr clouds reside on exactly those parts of the orbit where the line-of-sight components of the velocities are suppressed even further relative to the orbital motion. These effects explain why our model does a very good job reproducing the position of the star-forming ring, but is less successful at reproducing the line of sight velocities. It also serves as a warning against the limitations of our axisymmetric assumption: our model is capable of predicting at which galactocentric radii the star formation should occur, and how it should be regulated, but is not adequate to reproducing the detailed kinematics, which likely vary substantially in time in any event.

\begin{figure}
\includegraphics[width=\columnwidth]{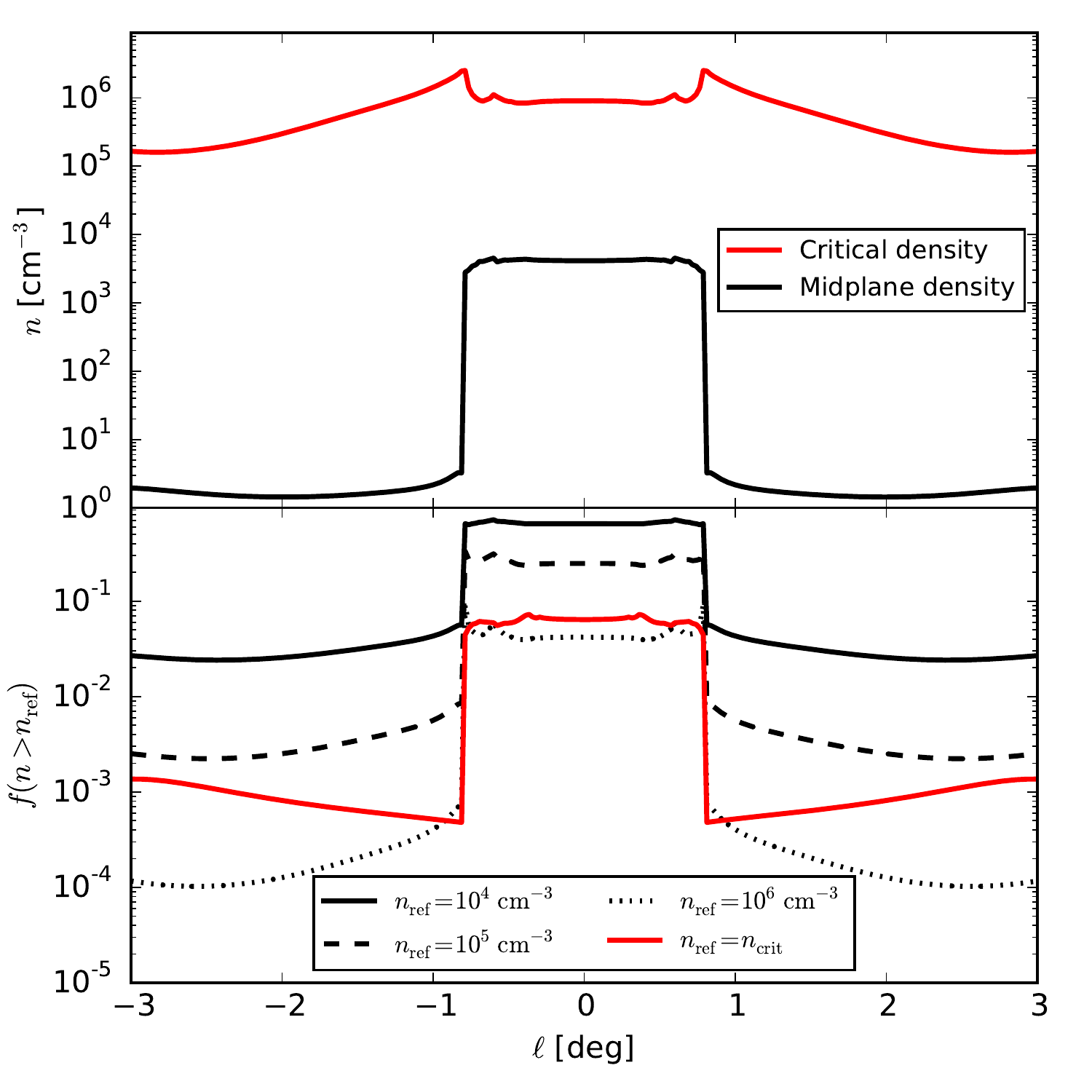}
\caption{
\label{fig:critdens}
Density structure of the cold interstellar medium for the same snapshot as shown in \autoref{fig:galmap}. Top panel: midplane volume density (black) and critical volume density threshold for star formation (red) as a function of Galactic longitude $\ell$. Bottom panel: mass fraction of all gas along the line of sight above a volume density $n_{\rm ref}$ as a function of Galactic longitude. The (solid, dashed, dotted) black lines represent $n_{\rm ref}=(10^4, 10^5, 10^6)$~cm$^{-3}$ and the solid red line indicates the gas mass fraction eligible for star formation, i.e., $n_{\rm ref}=n_{\rm crit}$, where $n_{\rm crit}$ is the critical density from the top panel. Throughout most of the Galactic Centre, the star-forming gas fraction is minor ($\ll1$ per cent), but it is predicted to be as high as several per cent in the star-forming ring.
}
\end{figure}

To facilitate future observational tests, we now present a number of simple predictions for the volume density structure of the cold interstellar medium in the Milky Way's CMZ. As demonstrated in \citetalias{krumholz15d}, our model predicts a strong increase towards the Galactic Centre of the midplane volume density ($\rho=\Sigma/2H_g$), the dense gas fraction, and the critical density threshold above which gas decouples from the turbulent flow and can collapse to form stars ($\rho_{\rm crit}=A\alpha_{\rm vir}{\cal M}^2\rho$, with $A$ a constant of order unity, see \citealt{krumholz05c,hennebelle08b, padoan11a, federrath12a}). We quantify this prediction by considering the gas in the midplane and assuming that it follows a lognormal volume density probability distribution function (PDF) as expected for an isothermal, supersonically turbulent medium \citep[e.g.][]{vazquez-semadeni94a,padoan97a,krumholz05c}. For simplicity, we adopt a mixture of compressive and solenoidal turbulence driving \citep[cf.][]{federrath12a} and assume that the magnetic field is not dynamically important. At each galactocentric radius, we calculate the density PDF from the midplane volume density and Mach number provided by our model, and also derive the critical density for star formation as in \citet{krumholz05c}. The PDFs are then used to determine the gas mass fractions above several different volume density thresholds as a function of Galactic longitude, where the integration along the line of sight is carried out by weighting each element by its local surface density.

The results of the above calculation are shown in \autoref{fig:critdens} for the same model snapshot as in \autoref{fig:galmap} (at $t=485$~Myr).\footnote{Across the star formation cycle in the star-forming ring (\autoref{fig:obs_cycle}), the dense gas fractions only vary by a factor of $\sim3$, whereas the critical density and midplane densities vary by factors of $\sim2$ and $\sim10$, respectively.} The top panel demonstrates that the midplane density in the gravitationally unstable, star-forming ring ($|\ell|\la1^\circ$) is much higher than elsewhere in the CMZ. The midplane densities of $n=10^3 - 10^4~{\rm cm}^{-3}$ provide a good match to the mean densities observed in the Galactic star-forming ring \citep[e.g.][]{bally10a,longmore12a,rathborne14b}. In addition, the critical density for star formation (which manifests itself in observed density PDFs as a power law deviation from the lognormal shape at high densities) is predicted to range from $n_{\rm crit}\sim10^5 - 3\times10^6~{\rm cm}^{-3}$ throughout the CMZ, with the highest values being reached in the star-forming ring. Such critical densities are orders of magnitude higher than those predicted for solar neighbourhood clouds, but are expected at the high pressures and densities of the CMZ gas \citep{kruijssen14b}. In gravitationally unstable gas, we predict $n_{\rm crit}\sim3\times10^6~{\rm cm}^{-3}$. This is remarkably consistent with the ALMA observations of the CMZ cloud G0.253+0.016 by \citet{rathborne14a}, who identify a power law deviation from the lognormal {\it column} density PDF that corresponds to a volume density of $n>10^6~{\rm cm}^{-3}$ when assuming spherical symmetry. The location of the high-density gas coincides with the only known site of star formation within the cloud (as traced by a water maser, see \citealt{lis94a}), providing further support to the interpretation that cloud-scale star formation in the CMZ is in accordance with the model presented here.

The bottom panel of \autoref{fig:critdens} quantifies the increase of the dense gas fraction towards the Galactic Centre. For different definitions of ``dense" ($n_{\rm ref}$, see the legend), the figure shows that the star-forming ring holds the highest dense gas fractions in the CMZ, ranging from several per cent (for $n_{\rm ref}=10^6~{\rm cm}^{-3}$) to nearly 100 per cent (for $n_{\rm ref}\sim10^4~{\rm cm}^{-3}$). The gas fraction eligible for star formation, i.e., $f(n>n_{\rm crit})$, is shown by the red line, and explains why star formation in the Milky Way's CMZ is mostly confined to $|\ell|<1^\circ$. Only at those longitudes does a non-negligible fraction of the gas reside at densities high enough to decouple from the turbulent flow and collapse to form stars.

Currently available observations confirm the prediction of our model that the majority of the cold interstellar medium in the CMZ resides at densities $n>10^4~{\rm cm}^{-3}$ \citep{longmore13a}, but observational tests of our predicted gas fraction above higher densities cannot be carried out yet, because no high-spatial resolution survey of the entire CMZ has been published. The few cases for which high-resolution observations are available match the predictions of \autoref{fig:critdens} \citep{kauffmann13a,rathborne14a}. However, a definitive test of our model requires a wide-field survey at arcsecond ($\sim0.08$ pc) resolution to enable the systematic mapping of the high-density gas and protostellar core population in the CMZ. This will be one of the main goals of the ongoing CMZoom Survey with the Submillimeter Array (SMA, PIs Keto \& Battersby), which is expected to reach densities of several $10^5~{\rm cm}^{-3}$. Future surveys with the Atacama Large Millimeter/submillimeter Array (ALMA) would grant access to even higher densities.

\subsection{Galactic Centre Star Formation Beyond the Milky Way}

Thus far we have focused on the Milky Way's CMZ, since that is the region for which we have the best measurement of the rotation curve and of the properties of the bar. However, there is every reason to believe that the Milky Way's centre is similar to that of other barred spiral galaxies, and thus that the phenomena we have investigated here should be generic in such systems. Inflows and bursts should occur in any CMZ where there is a bar to drive transport, a low shear region to trap the gas, and where the dynamical time at the low-shear region is shorter than the lifetimes of massive stars, preventing supernovae from establishing a time-steady equilibrium between driving and dissipation. What will other galaxies' CMZs look like if we observe them? To answer this question, we imagine observing the centre of an external galaxy and placing it on a Kennicutt-Schmidt (KS) plot, whereby we place the star formation rate per unit area on the $y$ axis and either the gas surface density $\Sigma$ or the gas surface density normalised by the orbital period $\Sigma/t_{\rm orb}$ on the $x$ axis.

Because such an exercise is necessarily resolution-dependent, we perform it in two ways. First, we consider an aperture of 750 pc centred on the (generic) galactic centre. Our motivation for this choice of size is that it is typical for large-scale nearby galaxy surveys such as THINGS \citep{walter08a, bigiel08a, leroy08a} and HERACLES \citep{leroy09a, leroy13a}. We compute the total gas mass and total star formation rate for all radial bins whose centres lie within the aperture, and obtain the area-normalised quantities by dividing both the gas mass and star formation rate by the total area of this region. For the orbital period, we use the value for the outermost bin within the 750 pc aperture.

Second, we consider a much higher resolution observation focused on the star-forming ring. For this case we identify the radius $r_{\rm peak}$ which has the highest time-averaged star formation rate, and we consider the ring of material at $r_{\rm peak}\pm 10$ pc. We use the gas mass, star formation rate, and area only of this region, and the orbital period at its outer edge. This produces an observation that is narrowly focused on the region of maximum star formation.

In both cases we use the observationally-inferred rather than true star formation rate in our computation. Strictly speaking the timescales for our ``observationally-inferred" star formation rate are appropriate only for an observation based on an ionised gas tracer, whereas many galactic centre observations use other tracers such as infrared. However, the timescales for ionisation and bolometric luminosity (which is what is closest to infrared) are not so disparate that we worry about this detail. We perform this exercise for runs m01r10f10, m03r10f10, and m10r10f10, which use our fiducial parameters for star formation and feedback, and vary in their mass accretion rates, as we would expect for a realistic population of galaxies with different bar strengths and gas contents.

\begin{figure*}
\includegraphics[width=\textwidth]{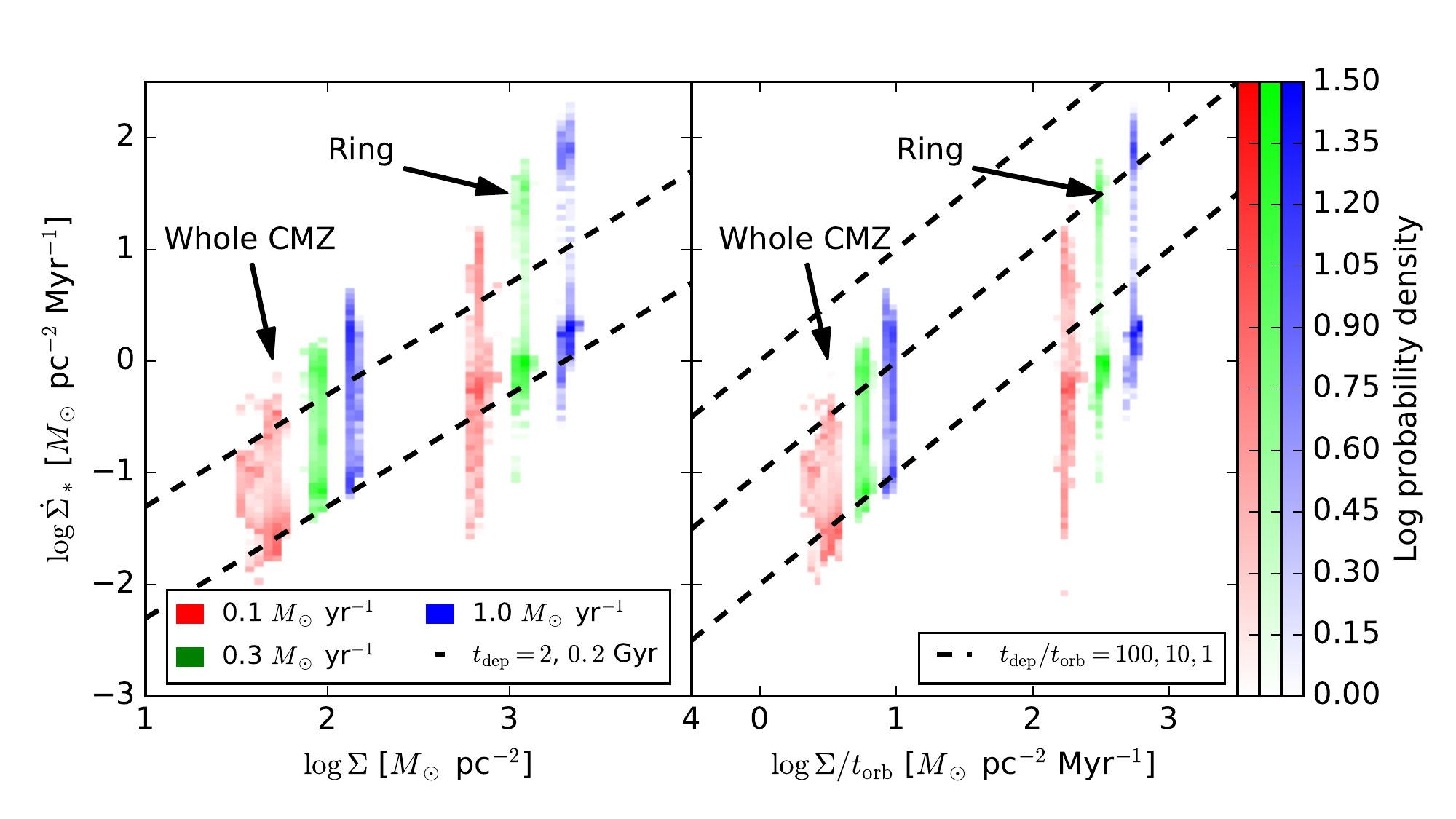}
\caption{
\label{fig:ksplot}
The observable properties of simulations m01r050f10 (red), m03r050f10 (green), and m10r050f10 (blue), corresponding to gas accretion rates $\dot{M}_{\rm in} = 0.1$, $0.3$, and $1.0$ $M_\odot$ yr$^{-1}$, in a Kennicutt-Schmidt plot. The left panel shows star formation rate per unit area $\dot{\Sigma}_*$ versus gas surface density $\Sigma$, while the right shows $\dot{\Sigma}_*$ versus surface density divided by orbital period, $\Sigma/t_{\rm orb}$. In each panel, colours indicate the log of the probability that the system would fall into the indicated pixel if observed at a random time $>200$ Myr of evolution in the simulation. We show the results both for an observation of the whole CMZ, and for one focusing on the ring of peak star formation, as indicated by arrows. For comparison, in the left panel the two black dashed lines show constant depletion times of 2 Gyr (bottom) and 200 Myr (top), respectively, while in the right panel they show gas depletion times of 100, 10, and 1 times the orbital period (bottom to top). See main text for details on how all quantities are computed.
}
\end{figure*}

We show the results in \autoref{fig:ksplot}. Examining the left panel, it is clear that the simulations with different accretion rates form a sequence that slides from the bottom left to the upper right of the KS plot. The lower extent of the locus of points occupied by a galaxy with a particular accretion rate, observed at 750 pc resolution, moves along a line of constant, $\sim 2$ Gyr depletion time, while the upper extent rises $\sim 1.5$ orders of magnitude higher in star formation rate above this.\footnote{We caution at this point that the extent of the vertical rise may be overestimated somewhat, because our axisymmetric model forces star formation events to be perfectly synchronised in azimuth, whereas in reality they are not. The main effect of this will be to compress the range of the points along the $\dot{\Sigma}_*$ axis somewhat, though probably more at the high end than the low end.} The points corresponding to a high resolution observation show similar qualitative behaviour to the low resolution ones, but with much larger scatter. However, in both cases galaxies spend roughly half their time near the line of 2 Gyr depletion time that characterises star formation in spiral galaxies at larger galactocentric radii, and about half their time scattered above this line, with a slight bias to being found at lower depletion time. These statistics are in very good agreement with the observed sample of \citet{leroy13a}.

The right panel of \autoref{fig:ksplot} tells a somewhat similar story. Measured at low resolution, galaxies spend about half their lives looking like their centres deplete on timescales of $\sim 100$ orbits, with this number dropping as low as $\sim 10$ orbits for $\sim 50\%$ of the time. Focusing on the ring where gas accumulates, star formation actually looks significantly \textit{less} efficient than on larger scales when measured in terms of the orbital period, with star formation rates rising to push the depletion time below 100 orbital periods only during outburst. This is simply a reflection of the fact that, during the quiescent period, $\epsilon_{\rm ff}$ is somewhat less than 1\% because the gas is supervirial, and the free-fall time is somewhat longer than the orbital period because the gas is not quite self-gravitating. Only when the gas becomes roughy virial and an outburst begins do we begin to have depletion times that approach $\sim 10$ orbital periods. This effect is not seen in the larger scale observations because, although the gas and star formation are all concentrated in a ring at $\sim 100$ pc, the orbital period being used is that measured at much larger galactocentric radii.

\subsection{Properties of the Cool Wind}
\label{ssec:windprop}

We next examine in more detail the properties of the galactic centre winds that are launched in our simulations. In particular, we are interested in the kinematics of the cold gas that is launched from the winds, as well as the properties of any hot, escaping gas and non-thermal particles. Observations constrain all of these quantities in the Milky Way.

First consider the cold gas driven upward by momentum injection. To obtain its velocity distribution, we again turn to the \citet{thompson16a} momentum-driven wind model. The central idea in this model is to consider a region of a galactic disc with mean surface density $\Sigma$, but where there are a wide range of local surface densities $\Sigma'$ as a result of turbulence. We then consider the vertical equation of motion for a particular fluid element near the top of the disc (i.e., at $z\sim H$) in a region with local surface density $\Sigma'$. This is
\begin{equation}
\frac{dv}{dt} \approx -g_{\rm gas} - g_* + \frac{d\dot{p}/dA}{\Sigma'}.
\end{equation}
Here $v$ is the vertical velocity, the first and second terms represent the force per unit mass exerted by gas and stars, and the final term represents the force per unit mass on the fluid element due to momentum injection from stellar feedback, which is provided at a rate per unit area $d\dot{p}/dA$. Note that, because gravity is a long-range force that is produced by material over a large area, the gravitational acceleration depends on the mean surface density $\Sigma$, while the acceleration due to feedback depends on the local one $\Sigma'$. Using our definitions of the Eddington injection rate, \autoref{eq:pdotedd}, and its non-dimensionalisation $x_{\rm crit}$, \autoref{eq:xcrit}, and defining $x=\ln \Sigma'/\Sigma$ for convenience, we can rewrite the equation of motion for a local fluid parcel as
\begin{equation}
\frac{dv}{dt} = (g_{\rm gas} + g_*) \left(e^{x_{\rm crit}-x}-1\right).
\end{equation}
Gas is ejected in regions where the local surface density is low enough that $x<x_{\rm crit}$, and thus the left-hand side is positive, indicating an upward acceleration. If the gas is accelerated over a distance $\sim r$, and $x$ remains constant as this happens, then its final speed will be
\begin{equation}
\label{eq:vx}
v \approx v_{\rm esc} \sqrt{\left(e^{x_{\rm crit}-x}-1\right)},
\end{equation}
where we may think of $v_{\rm esc} = \sqrt{2r(g_{\rm gas}+g_*)}$ as the characteristic escape speed for gas flowing out in the wind. Since this provides a mapping between the local surface density $\Sigma'$ and the outflow velocity, we can obtain the distribution of outflow velocity by combining this mapping with the distribution of surface densities $dm/dx$ produced by turbulence. Specifically, if we let $u = v/v_{\rm esc}$, then for any given $u$ we can invert \autoref{eq:vx} to obtain $x(u)$, and we can write
\begin{equation}
\frac{dm}{du} \propto \left|\frac{dx}{du}\right| \left(\frac{dm}{dx}\right)_{x=x(u)} \propto \frac{2u}{u^2+1}\left(\frac{dm}{dx}\right)_{x=x(u)}.
\end{equation}
Following \citet{thompson16a}, the mass distribution $dm/dx$ is a lognormal given by
\begin{equation}
\frac{dm}{dx} = \frac{1}{\sqrt{2\pi\sigma_x^2}} \exp\left[-\frac{(x-\sigma_x^2/2)^2}{2\sigma_x^2}\right],
\end{equation}
where we compute $\sigma_x$ from the Mach number as outlined in \citeauthor{thompson16a}. Armed with these relationships, we can compute the velocity distribution of outflowing gas from each computational zone, and by summing over zones we can obtain the full velocity distribution at every instant.

\begin{figure}
\includegraphics[width=\columnwidth]{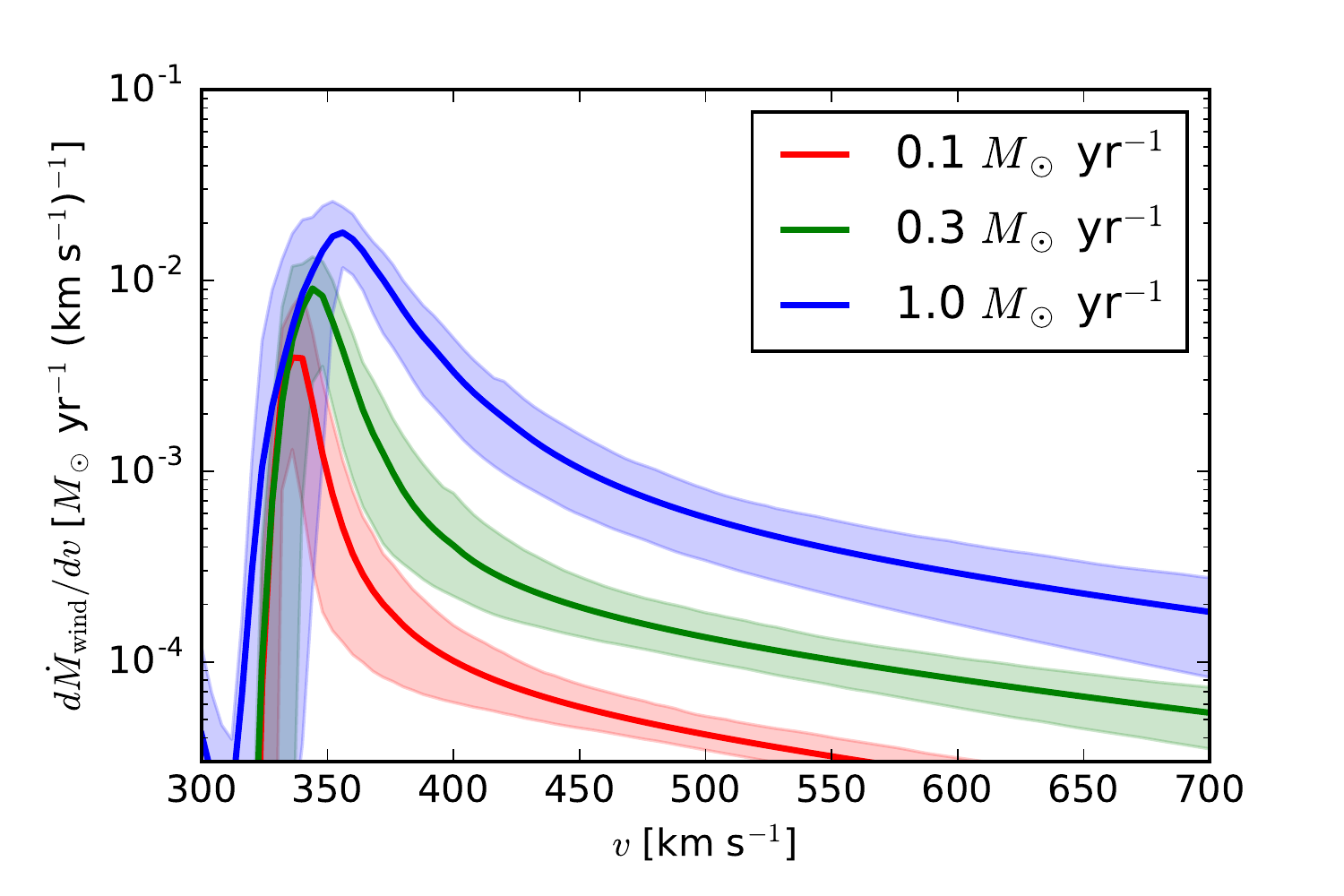}
\caption{
\label{fig:wind}
Wind velocity distribution $d\dot{M}_{\rm wind}/dv$ in simulations m01r050f10, m03r050f10, and m10r050f10, corresponding to accretion rates of $\dot{M} = 0.1$, 0.3, and 1.0 $M_\odot$ yr$^{-1}$, as indicated. Solid lines indicate the mean over all times $>200$ Myr, while shaded regions indicate the range from 10th to 90th percentile in time.
}
\end{figure}

We perform this computation for runs m01r10f10, m03r10f10, and m10r10f10 and plot the resulting wind velocity distribution in \autoref{fig:wind}. We see that the wind velocity distribution strongly peaks at $\approx 350$ km s$^{-1}$, which is the escape speed from the star-forming ring. There is a tail to higher velocities, which becomes increasingly prominent at higher star formation rates, but the great majority of the mass emerges close to the escape speed. The wind launch rate at a given velocity varies by a factor of $\sim 3-5$ at any given time.

\subsection{The Hot Wind and Non-Thermal Particles}

Only a small fraction of the total energy injected by supernovae goes into driving either turbulent motions or the cool wind. Indeed, one can see this immediately from a simple argument. In a region with a steady star formation rate per unit area $\dot{\Sigma}_*$, if we have one supernova per mass $M_{\rm SN}$ of stars formed, then the supernova rate per unit area is $\dot{\Sigma}_*/M_{\rm SN}$. The momentum and energy injected per supernova are $p_{\rm SN}$ and $E_{\rm SN}$, respectively, giving momentum and energy injection rates per unit area $\dot{p}_{\rm SN} = p_{\rm SN} \dot{\Sigma}_*/M_{\rm SN}$ and $\dot{E}_{\rm SN} = E_{\rm SN} \dot{\Sigma}_*/M_{\rm SN}$. The energy injection rate into turbulent motions is (\autoref{eq:edotturb})
\begin{equation}
\left(\frac{d\dot{E}}{dA}\right)_{\rm SF,turb} = \sigma \frac{p_{\rm SN}}{M_{\rm SN}}\dot{\Sigma}_*,
\end{equation}
and the ratio of this to the total supernova energy budget is
\begin{equation}
\frac{(d\dot{E}/dA)_{\rm SF,turb}}{(d\dot{E}/dA)_{\rm SN}} \approx \sigma \frac{p_{\rm SN}}{E_{\rm SN}} \approx \frac{\sigma}{170\,\mathrm{km\, s}^{-1}},
\end{equation}
where the numerical evaluation is for our canonical values $p_{\rm SN} = 3\times 10^5$ $M_\odot$ km s$^{-1}$ and $E_{\rm SN} = 10^{51}$ erg. Thus for the values of $\sigma$ found in our simulation, only a small portion of the supernova energy budget is consumed by driving turbulence. Similarly, the wind kinetic luminosity per unit area is of order
\begin{equation}
\left(\frac{d\dot{E}}{dA}\right)_{\rm wind,kin} \approx \frac{1}{2}\dot{\Sigma}_{\rm wind} v_{\rm esc}^2,
\end{equation}
and the ratio of this to the supernova energy injection rate is
\begin{equation}
\frac{(d\dot{E}/dA)_{\rm wind,kin}}{(d\dot{E}/dA)_{\rm SN}}
= \frac{\dot{\Sigma}_{\rm wind}}{\dot{\Sigma}_*} \left(\frac{M_{\rm SN} v_{\rm esc}^2}{2E_{\rm SN}}\right) \approx 0.09 \eta,
\end{equation}
where the numerical evaluation is for $M_{\rm SN}=100$ $M_\odot$ and $v_{\rm esc} = 300$ km s$^{-1}$, and $\eta = \dot{\Sigma}_{\rm wind}/\dot{\Sigma}_*$ is the mass loading factor, which for our simulations is $\lesssim 1$.

Thus we conclude that neither launching the wind nor driving the turbulence consumes an appreciable fraction of the total supernova energy available. Instead, the energy released by supernovae must either be lost to radiation, or must go into a hot wind that carries it out. Unfortunately our model does not allow computation of the partition between these two forms of energy loss, and the simulations that have been published to date are not helpful in addressing this question -- answering it correctly requires simulating with enough resolution to resolve the Sedov-Taylor phase of supernova remnant expansion, without using any artificial methods to lower the density in the vicinity of the supernovae. As we discuss in \autoref{ssec:comparison}, no published simulations meet these criteria. Observations of superbubbles away from galactic centres suggest that radiation cannot be the primary loss mechanism \citep{rosen14a}, but it is unclear whether we can generalise these conclusions to the very different environment of the CMZ. If radiative losses do not dominate, then the hot wind must carry an energy flux of order $\dot{M}_* E_{\rm SN}/M_{\rm SN}$, and given our result that $\sim 50\%$ of the incoming gas is converted to stars, this implies an energy flux $\dot{E}_{\rm hot} \sim \dot{M}_{\rm in} E_{\rm SN}/2 M_{\rm SN} \approx 10^{41} (\dot{M}_{\rm in}/M_\odot\,\mbox{yr}^{-1})$ erg s$^{-1}$. This material will mostly be launched from the star-forming ring at 100 pc.

The non-thermal particle energy injection rate should be $\sim 10\%$ of the hot gas energy budget, implying $\dot{E}_{\rm non-therm} \sim 10^{40} (\dot{M}_{\rm in}/M_\odot\,\mbox{yr}^{-1})$ erg s$^{-1}$. Because non-thermal particle acceleration happens even if the hot gas does not vent, and because the non-thermal particles have long mean-free paths and thus should be able to escape the disc, this estimate should hold even if the primary loss mechanism is radiation rather than a hot wind. Cosmic ray escape is confirmed observationally \citep[][and references therein]{crocker12a}: the diffuse non-thermal emission from the CMZ in radio continuum and $\gamma$-ray bands implies that only a small fraction ($\lesssim 10$\%) of the power injected into non-thermal particles accelerated in the CMZ is lost {\it in situ} radiatively; most of this power is carried off by escaping cosmic rays (and is claimed by \citet{crocker15a} to be radiated on much larger size scales in the Fermi Bubbles). This is also consistent with the upper limit on the dense gas ionisation rate by cosmic rays ($\zeta_{\rm CR} < 10^{-14}$ s$^{-1}$) that is implied by the observed temperature distribution of formaldehyde, which extends down to gas temperatures as low as $T \sim 40$ K and exhibits substantial cloud-to-cloud variation, showing that cosmic rays do not set the gas temperatures of CMZ clouds \citep{ginsburg16a}.
These findings, in concert with the measured, hard spectrum of the diffuse, non-thermal radio continuum and $\gamma$-ray radiation from the CMZ, supports the notion that the region's cosmic ray population is advected away with the putative hot outflow.

\subsection{Relationship to Other Models}
\label{ssec:comparison}

A number of other authors have proposed models of the CMZ, and it is worth commenting on the ways in which the model we propose here compares to theirs. \citet{kim11b} conduct 3D SPH simulations of the flow of gas in a barred spiral potential chosen to represent the Milky Way, including star formation and feedback, and find that the gas forms a nuclear ring $\sim 200$ pc from the Galactic Centre, somewhat further out than we observed in the Milky Way. The somewhat different location of the ring in their simulations is likely a result of the potential they adopt, which is a simplified model that, aside from the bar perturbation, possesses uniform shear, rather than having a low-shear region as our more realistic potential does. Once the simulation reaches equilibrium, \citeauthor{kim11b} find that the gas mass in the CMZ is about constant at $\sim 10^7$ $M_\odot$, and the star formation rate is relatively steady at $\sim 0.05$ $M_\odot$ yr$^{-1}$. While these figures are quite similar to the averages of our fiducial case, \citeauthor{kim11b}'s simulation does not show the bursty behaviour we observe in our model. In contrast, we use the measured potential, which does possess a shear minimum at the observed location of the gas ring. The lack of burstiness in their star formation rate is likely dependent on their feedback implementation (which is described in \citealt{saitoh08a}), and differs from what some other simulations find. We discuss this topic further below.

\citet{kim12b}, \citet{kim12c} and \citet{li15b} perform high resolution 2D simulations of gas flows in the presence of bars with a wide range of parameters, and find that, for sufficiently slowly-rotating bars, the typical outcome is a ring in an $x_2$ orbit at distances of hundreds of pc from the centres of their simulated galaxies; the inner regions of their rings do correspond roughly to where the rotation curve turns over to solid body, consistent with the mechanism for ring formation that we have proposed. They do not include self-gravity, star formation, or feedback in their simulations, and thus do not make any predictions regarding these phenomena.

\citet{crocker12a} and \citet{crocker15a} provide a one-zone model for the CMZ, focused on reproducing the properties of the outflow and non-thermal emission found there. Because the model is steady-state and one-zone, it does not address the questions of spatial and temporal variation on which we focus. Conversely, however, our model does not address the properties of the outflow or the non-thermal emission, and it would therefore be extremely interesting to extend our model using \citeauthor{crocker15a}'s machinery for the treatment of non-thermal emission. We plan to do so in future work.

Most recently, \citet{torrey16a} presented a model and a set of 3D simulations for the behaviour of star formation in the central regions of galaxies. Their main finding is that star formation in these regions is bursty, with burst timescales of $\sim 50$ Myr. The mechanism they identify is similar to what we find in our simulations, namely that the  dynamical time is comparable to the time for which supernovae go off after a starburst, so star formation feedback tends to ``overshoot", leading to alternating cycle of starburst and quenching rather than a steady state. The $\sim 50$ Myr variability timescale they find is a bit longer than ours, likely because they do not include a non-axisymmetric stellar potential that is capable of driving mass inflows. As a result, only the non-axisymmetric self-gravity of the gas and the galactic fountain are available as mechanisms to refill the gas in the CMZ once it has been expelled. In contrast, the outer parts of our simulated disc continue to move mass inward efficiently as a result of bar-driven instabilities regardless of what is happening in the star-forming region. This difference probably causes the longer delay in restarting star formation in their simulations compared to ours.

A further difference between \citet{torrey16a}'s model and ours is that, in their simulations, the mechanism responsible for causing bursts is gas expulsion rather than changes in the depletion time of the gas, in contrast to our model where the opposite holds. It is unclear which result is more realistic. While their simulations of course are 3D rather than 1D, the results are also in strong contrast to those obtained by \citet{kim11b} who also use 3D simulations, and much of the result appears to depend on the sub-grid models used for feedback. 
Neither \citet{kim11b} nor \citet{torrey16a} resolve the Sedov-Taylor phase of supernova blast waves, and as a result they are forced to rely on approximate models for supernovae to avoid the ``overcooling" problem \citep{katz92a}, whereby simulations that do not resolve blast waves overestimate the rate of radiative losses from supernova remnants. \citet{kim11b} handle this problem using decoupled wind particles, following the prescription of \citet{okamoto08a}, while \citet{torrey16a} directly add radial momentum to the gas in cases where they do not resolve the blast wave. Although our model is also based on momentum injection, it differs from \citeauthor{torrey16a}'s approach in that we explicitly model the interaction of this momentum with the density structure of the turbulent medium, and determine the wind mass flux based on this model. In contrast, \citeauthor{torrey16a}'s simulations do not resolve turbulence at the momentum injection scale (since momentum is injected at the resolution scale), and thus their approach implicitly differs from ours. 
None of these approaches are perfect, and to the extent that the nature of the starbursts depends on them, the results of any model are somewhat suspect.

\section{Conclusion}
\label{sec:conclusion}

In this paper we present a simple dynamical model for star formation in the nuclear regions of galaxies. We focus on the Milky Way CMZ, since this is the only nuclear region for which we have available a very high resolution measurement of the rotation curve, but we argue that the phenomena we find there should be generic in barred spiral galaxies. This model captures several essential elements that combine to produce the distinctive behaviour of star formation in these regions; some of these elements have been explored before, but the model we present here is the first to combine them all. These elements are as follows.

\textit{Mass Transport by Acoustic Instability.} The nuclear regions of galaxies have gas depletion times far smaller than the Hubble time, so for them to continue star formation at the present epoch requires a constant resupply of mass. At large galactocentric radii, the required transport is likely provided by gravitational instability \citep{krumholz10c, forbes12a, forbes14a, goldbaum15a, goldbaum16a, schmidt16a}, but this mechanism is suppressed in nuclear regions by strong shear. However, inside the inner Lindblad resonance of the bar, another transport mechanism becomes available: acoustic instability driven by the bar, which thrives in regions of high shear (\citealt{montenegro99a}, \citetalias{krumholz15d}). This instability both moves gas inward and drives turbulence, keeping it gravitationally stable and suppressing star formation as the gas is transported. This explains the paucity of star formation in the Milky Way found at radii from $\sim 150 - 500$ pc.

\textit{The Effects of the Rotation Curve.} Because the mechanism for mass transport and turbulence driving is sensitive to the amount of shear, it must cease where the rotation curve switches from flat to (near-)solid body, which is a common feature of galactic centres. This causes gas to accumulate and become gravitationally unstable in a particular region. Thus nuclear star formation is characterised by the presence of persistent, long-lived, ring-like structures, rather than by transient molecular clouds arranged in either grand design or flocculent spiral patterns. In the Milky Way, this structure is found at $\approx 100$ pc from the Galactic Centre, and manifests as a partially-filled ring, within which the bulk of the CMZ's dense gas and young star clusters reside.

\textit{Evolutionary state of the Milky Way's CMZ.} Based on a detailed comparison of our model to the observed properties of the CMZ, we predict that the star-forming ring currently resides at a star formation minimum, with the previous starburst having taken place 8 Myr ago. In the context of our model, the Arches and Quintuplet clusters represent the final clusters to have formed during this latest starburst ($\sim5$~Myr ago). By contrast, the CMZ ``dust ridge" (spanning in projection from Sgr A$^*$ to Sgr B2 and containing the most massive and densest molecular clouds in the CMZ) will collapse and form stars first during the onset of the upcoming starburst (expected in 1--2 Myr). We also provide quantitative predictions for the dense gas fraction and critical density for star formation as a function of Galactic longitude, finding that dense ($n>10^{(4,5,6)}~{\rm cm}^{-3}$) gas and star formation are mostly confined to $|\ell|<1^\circ$ (or $R<150$ pc). This matches the position of the 100-pc stream in the CMZ \citep{molinari11a,kruijssen15a}, as well as the major known sites of recent star formation, such as Sgr B2, Sgr C, and the Arches and Quintuplet clusters.

\textit{Supernova Feedback-Regulated Star Formation.} Within the ring-like structure acoustic instability is unable to drive turbulence or transport mass, and thus the gas is liable to become gravitationally unstable and begin vigorous star formation. When a starburst begins, there is initially little feedback, because supernovae, which provide the most important feedback mechanism, are delayed from $4-40$ Myr after the onset of star formation. This leads to an overshoot, so that, when supernovae do begin to occur, the system does not settle into forming stars at a steady state. Instead, the supernovae raise the velocity dispersion, scale height, and virial parameter in the star-forming ring so that the star formation rate falls dramatically. Star formation remains suppressed until there is time for supernova feedback to taper off and for turbulence to decay, leading to the resumption of star formation. Because this cycle occurs within a coherent star-forming structure whose location is fixed by the galactic rotation curve,  the overall nuclear star formation rate and depletion time undergo large oscillations. In the Kennicutt-Schmidt diagram, which measures the gas depletion time, this results in nuclear regions undergoing large excursions, with some appearing similar to ``normal" galaxies and others resembling starburst galaxies.

\textit{Supernova-Driven Winds.} The supernovae that regulate star formation also drive a two-phase wind off the star-forming ring. The cool phase of the wind dominates the mass flux, and carries off mass at a rate that is comparable to or slightly smaller than the mass flux going into stars. However, it carries away relatively little of the supernova energy budget. The energy is most likely carried by a hot phase that accompanies the cool, momentum-driven wind, though it could conceivably also be lost to radiation. Some of this energy likely goes into the production of non-thermal particles as well. Wind launching is bursty like star formation, but the magnitude of the variation is somewhat smaller than that of the star formation rate.

Taken together, these elements are able to explain the observed properties of nuclear star formation in Milky Way-like galaxies in general, and in the Central Molecular Zone of the Milky Way in particular.

\section*{Acknowledgements}

We thank the referee for helpful comments. MRK acknowledges support from an Australian Research Council Discovery Project (DP160100695). JMDK gratefully acknowledges financial support in the form of a Gliese Fellowship and an Emmy Noether Research Group from the Deutsche Forschungsgemeinschaft (DFG), grant number KR4801/1-1. RMC is the recipient of an Australian Research Council Future Fellowship (FT110100108). MRK and JMDK thank the Aspen Center for Physics, which is supported by NSF Grant PHY-1066293, for its hospitality during the early phases of this work.

\small
\bibliographystyle{mn2e}
\bibliography{refs}

\label{lastpage}

\end{document}